\documentclass[12pt,preprint]{aastex}
\usepackage{color}\newcommand{\del}[2]%
{\frac{\mathrm{d}{#2}}{\mathrm{d}{#1}}}
\newcommand{\Del}[2]%
{\frac{\mathrm{D}{#2}}{\mathrm{D}{#1}}}
\newcommand{\ddel}[2]%
{\frac{\mathrm{d}^2{#2}}{\mathrm{d}{#1}^2}}
\newcommand{\pdel}[2]%
{\frac{\partial{#2}}{\partial{#1}}}
\newcommand{\pddel}[2]%
{\frac{\partial^2{#2}}{\partial{#1}^2}}

\newcommand{\Ms}{M_{\odot}}
\newcommand{\km}{\,\, \mathrm{km}}

\newcommand{\gauss}{\,\, \mathrm{G}}

\newcommand{\gpcmc}{\,\, \mathrm{g \,\, cm^{-3}}}

\newcommand{\radps}{\,\, \mathrm{rad \,\, s^{-1}}}

\newcommand{\ergps}{\,\, \rm erg \,\, {s}^{-1}}

\slugcomment{}
\shorttitle{Gravitational Waves from Collapsars}

\begin{document}

\title{Gravitational Wave Signatures of Hyperaccreting Collapsar Disks}

\author{Kei Kotake\altaffilmark{1,2}, Tomoya Takiwaki\altaffilmark{2}, and 
 Seiji Harikae\altaffilmark{3}}
\affil{$^1$Division of Theoretical Astronomy, National Astronomical Observatory of Japan, 2-21-1, Osawa, Mitaka, Tokyo, 181-8588, Japan}
\email{kkotake@th.nao.ac.jp}
\affil{$^2$Center for Computational Astrophysics, National Astronomical Observatory of Japan, 2-21-1, Osawa, Mitaka, Tokyo, 181-8588, Japan}
\affil{$^3$Quants Research Department, Financial Engineering Division, Mitsubishi UFJ Morgan Stanley Securities Co., Ltd., Marunouchi Bldg., 2-4-1, Marunouchi, Chiyoda-ku, Tokyo, 100-6317}

\begin{abstract}
By performing two-dimensional special relativistic (SR) magnetohydrodynamic
 simulations, we study possible signatures of gravitational waves (GWs)  
 in the context of the collapsar model for long-duration gamma-ray bursts. 
 In our SR simulations, the central black hole is treated 
 as an absorbing boundary. By doing so, we focus on the GWs generated by 
 asphericities in neutrino
 emission and matter motions in the vicinity of the hyperaccreting disks.
 We compute nine models by adding initial angular momenta and magnetic fields parametrically to a precollapse core of a $35 M_{\odot}$ progenitor star.
 As for the microphysics,
 a realistic equation of state is employed and the 
neutrino cooling is taken into account via a multiflavor neutrino leakage scheme.
 To accurately estimate GWs produced by 
anisotropic neutrino emission, we perform a ray-tracing analysis in general relativity
 by a post-processing procedure.
 By employing a stress formula that includes
  contributions both from magnetic fields and special relativistic corrections,
 we study also the effects of magnetic fields on the gravitational waveforms. 
 We find that the GW amplitudes from
 anisotropic neutrino emission show a monotonic increase with time, whose 
 amplitudes are much larger than those from matter motions of the accreting material.
 We show that the increasing trend of the neutrino GWs stems from 
 the excess of neutrino emission in the direction near parallel to 
 the spin axis illuminated from the hyperaccreting disks. 
 We point out that a recently 
proposed future space-based interferometer like Fabry-Perot type DECIGO would permit the 
detection of these GW signals within $\approx$ 100 Mpc.
\end{abstract}

\section{Introduction}
Gamma-ray bursts (GRBs) are one of the most energetic phenomena
 in the universe. Thanks to {\it Swift} observations{\footnote {http://www.swift.psu.edu/}}, it has now become evident that GRBs are basically categorized into two, 
namely short-hard and long-soft bursts (e.g., \citet{hjorth11,nakar07} 
for recent reviews). 
More surprisingly, GRBs with some mixed features of the two types have been reported
(e.g., \citet{gehre06}, \citet{galy06}) possibly necessitating a new classification
 \citep{lue10}. The mystery of their central engines 
 seems to be thickening, which has long puzzled astrophysicists
 since the accidental discovery in the late 1960s
(see \citet{meszaros06} for review).
 Regarding the long-duration GRBs (LGRBs), 
 robust associations of the underlying supernovae with a handful of LGRBs
(e.g., \citet{gala98,hjorth03,stan03,malesani04,modjaz06,pian06},
 and collective references in \citet{woos_bloom,zhang11}) and the 
 fact that their host galaxies are typically irregular with intense
 star formation \citep{fruchter06} suggest that they 
 are likely related to the deaths of massive stars and the "collapsar" model
  has been widely recognized as the standard scenario for LGRBs
\citep{woos93,pacz98,macf99}.


 In the scenario, the collapsed iron core of a massive star forms a temporary disk
 around a few $M_{\odot}$ black hole (BH) and accretes at a high rate 
($\sim 0.1 - 10 M_{\odot}/{\rm s}$, e.g., \citet{popham,matteo02,kohri05,chen,zala} 
and references therein), whose gravitational binding energy is the 
 driving source of the central engine. \citet{paz90} and \citet{mezree} pioneeringly 
proposed that pairs of neutrino and anti-neutrino illuminated from the hyperaccreting 
disks that annihilate into electron and positron 
(e.g., $\nu + {\bar \nu} \rightarrow e^{-}+ e^{+}$, hereafter 
``neutrino pair annihilation'')  can supply sufficient energy to launch
 GRB outflows by heating material in the polar funnel regions.
 In addition, it is suggested that the strong magnetic fields in the
cores of order of $10^{15} \gauss$ play also an active role both for driving 
the magneto-driven jets and for extracting a significant amount of
energy from the central engine (e.g.,
\cite{bz77,mizu04a,mizu04b,mckinney,mcki07a,mcki07b,komi,komi07a,komi09,bark08,naga09},
 and 
references therein). 

Although various possibilities including magnetar models (e.g., 
\citet{dai98,thom04,uzde07a,bucci}) have been
 proposed so far, there has been no direct evidence to pin down the mechanism of 
the central engine.
 This is mainly because it is difficult to extract the information 
from conventional astronomy by electromagnetic waves, since   
 high-energy photons are absorbed through interactions in the source and 
by the photon backgrounds. Alternatively, 
gravitational waves (GWs) are expected to be a primary 
observable to decipher the mechanism of the engine, because they imprint a live 
information hidden deep inside the stellar core and they carry the information directly 
to us without being affected in propagating to the earth.

 Currently long-baseline laser interferometers such as
LIGO \citep{firstligonew},
VIRGO\footnote{http://www.ego-gw.it/},
GEO600\footnote{http://geo600.aei.mpg.de/},
and TAMA300 \citep{tamanew} are operational 
(see, e.g., \citet{hough} for a recent review).
For these detectors, core-collapse supernovae (CCSNe) have been proposed as one of 
the most plausible GW sources,
 therefore an extensive study of the GW predictions 
 based on sophisticated numerical modeling has been carried out so far 
(see, for example, \citet{ott_rev,fryer11,Kotake11_rev} for recent reviews).
 It is noted however that most of them have paid attention to CCSNe that leave behind 
neutron stars (NSs) after explosions. For a reliable prediction of GWs from CCSNe, 
one needs to perform multi-D hydrodynamic simulations equipped with a precise 
neutrino transport scheme which follows the dynamics starting from 
stellar core-collapse, core-bounce, through shock-stall and subsequent growth of 
hydrodynamic instabilities, the neutrino-driven shock revival, to stellar explosion
 in a consistent manner. This is one of the most challenging subjects in computational 
 astrophysics (e.g., \citet{jank07}). 

But the numerical modeling to test the collapsar scenario could be much more demanding.
 One needs to trace a new path that bifurcates from
 the above story after bounce, namely to the BH formation (phase 1), 
evolution of the surrounding accretion disk including energy deposition to the 
polar funnel region by neutrinos and/or magnetic fields (phase 2), 
to the launching of the fireballs (phase 3)\footnote{Here for convenience we call 
each stage as phase 1, 2, etc.}. 
 This apparently necessitates the multidimensional magnetohydrodynamic (MHD) simulations
 not only with general relativity (GR) for handling the BH formation, but also with the 
multi-angle neutrino transfer for treating highly anisotropic neutrino radiation 
from the disks. In the business of CCSN simulations, the most up-to-date
simulations\footnote{assisted by accelerating computer powers} can now follow the multi-angle neutrino 
transport but limited to a Newtonian case \citep{hubeny,ott_multi,sumi12},
 or handle GR with a sophisticated neutrino transport
 \citep{mueller} but not applicable to a rapidly rotating case (due to the assumption of 
 conformal flatness). 

  Various approximate approaches have been therefore undertaken in the business of the 
collapsar simulations. In the phase 1, GR simulations \citep{shib06}
   updated to implement a neutrino cooling have reported recently 
\citep{seki11}, in which the dynamics after the BH formation to the formation of 
accretion disk was first consistently followed. The numerical studies of the phase 
2 are concerned with the subsequent evolution of accretion disk and the outflow 
formation in the polar funnel region till the jets become mildly relativistic. 
The central BH has been traditionally treated by a fixed metric technique in GR 
simulations (e.g.,\citet{mizu04a,devi05,hawl06,mcki07b,komi07a,bark08}) or 
by an absorbing boundary in the Newtonian simulations
 (e.g., \citet{macf99,prog03c,fuji06,naga07,lopez09,camara} or special relativistic 
simulations (e.g., \citet{hari09}). As for the microphysics in
 these simulations, except for \citet{fuji06,naga07,hari09,shiba},
 a realistic nuclear equation of state (EOS)
 has been replaced by a very phenomenological one (like a gamma-law or polytrope) and the neutrino cooling (and heating) has been often
 neglected for simplicity. Numerical studies of the phase 3 are 
mainly concerned with the dynamics later on, namely, the jet
propagation to the breakout from the star, by assuming 
 a manual energy input to the polar funnel region (see, e.g., 
\citet{aloy00,zhang03,lazzati,nagakura,nagakura2} and references therein).
All of the studies mentioned above may be regarded as complimentary in the sense that 
the different epochs are focused on, with the different initial conditions for the 
numerical modeling being undertaken.

To the best of our knowledge, \citet{ott2011} is the only work which 
 extracted the GW signals based on their collapsar simulations (in 
 the phase 1). Based on their
 three-dimensional (3D) GR simulations of a 75 $M_{\odot}$ star with the use of 
a polytropic EOS, they pointed out that 
 the significant GW emission is associated at the moment of
 the black hole formation, which can be a promising target of the advanced LIGO for a 
Galactic source. In contrast to such a paucity of the GW predictions based on
 numerical simulations of collapsars, a number of 
semi-analytical estimates have been reported so far, which predict 
a significantly strong GW emission due to possible density-inhomogeneities 
\citep{mineshige}, 
bar or fragmentation instabilities in the collapsar's accretion torii
 (e.g., \citet{putten01,davies02,fryer02,kobayashi03,piro07,corsi09}, and
 collective references in \citet{fryer11}), and the precession 
 of the disks due to GR effects \citep{romero10,sun}. The predicted 
 GW amplitudes are typically high enough be visible to advanced-LIGO class detectors for
 a 100 Mpc distance scale, which is 
 about four orders-of-magnitudes larger than the numerical estimate 
at the black formation \citep{ott2011}.
 In addition to these GWs produced by non-spherical matter 
motions, \citet{hiramatsu,suwa} pointed out that anisotropic neutrino
 emission from accretion disk could be the source of 
 GWs from collapsars, which was originally proposed as an equally 
important GW source to the matter GW in the context of CCSNe \citep{epstein}.
Since these GWs from
 collapsars would be a smoking-gun signature of the central engine 
 in coincident with the conventional electromagnetic messengers as well 
as neutrinos\footnote{Especially for a nearby GRB source, e.g., \citet{ando_new}.},
 it will be very important to put forward theoretical predictions of 
  GW signals based on the collapsar simulations, as has been done in the 
  business of CCSNe.

In this work, we study possible GW signatures in the hyperaccreting 
 collapsar disks\footnote{which corresponds to evolution in the phase 2} by performing 
two-dimensional special relativistic (SR) MHD simulations 
 of accretion torii around a black hole. We compute nine models 
by adding angular momenta and magnetic fields parametrically 
 to a precollapse core of a $35 M_{\odot}$ progenitor star \citep{woos06}.
 As for the microphysics, a realistic equation of state is employed and the 
neutrino cooling is taken into account via a multiflavor neutrino leakage scheme.
 In our SR simulations, the central black hole is treated 
 as an absorbing boundary. By doing so, we focus on the GWs generated by 
 asphericities in neutrino emission and matter motions in the vicinity of the 
 accretion disks.
 To accurately estimate GWs produced by 
anisotropic neutrino emission, we perform a ray-tracing analysis
 in GR by a post-processing procedure.
 By employing a stress formula that includes
  contributions both from magnetic fields and special relativistic corrections,
 we study also the effects of magnetic fields on the gravitational waveforms. 
 Then we discuss their detectability by performing a spectrum analysis.

The paper opens up with descriptions of the initial models and numerical methods 
(section \ref{methods}). 
  The main results are given in Section \ref{results}.  
We summarize our results and discuss their implications in Section \ref{summary}.
\clearpage
\section{Numerical Methods and Models} \label{methods}
\subsection{Initial Models}

\begin{figure}[htbp]
\begin{center}
\epsscale{.7}
   \plotone{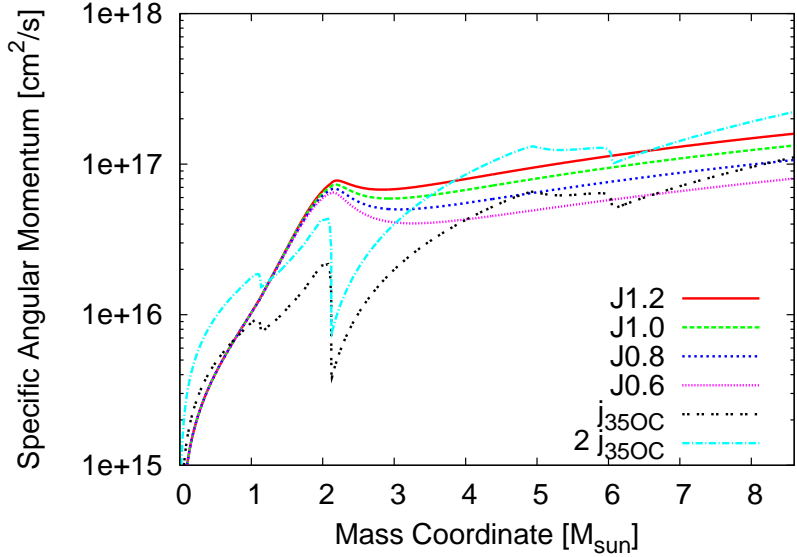}
\caption{Profiles of the specific angular momentum for model 35OC 
 (\citet{woos06}, labeled by $j_{35 \rm{OC}}$ and the one amplified 
 by a factor of 2 (labeled by $2 j_{35 \rm{OC}}$) and for 
models J0.6, J0.8, J1.0, and, J1.2 (from bottom
 to top), respectively.}
\label{f1}
\end{center}
\end{figure} 

Regarding our precollapse model, we take a 35 $_{\odot}$ star
 (model 35OC) in \citet{woos06} that is supposed to be one of the most promising
 GRB progenitor models.
 To study the effects of rotation systematically,
 we take the following rotational profiles as
\begin{equation}
 \Omega(r,\theta)=\frac{\Omega_0 X_0^4 + \alpha 
\Omega_\mathrm{lso}(M(X))X^4}{X_0^4+X^4}, 
 \label{eq:iniang}
\end{equation}
where $\alpha, \Omega_0,$ and $X_0$ are model parameters.
$M(X)$ is the mass coordinate at the cylindrical radius ($X=r\sin\theta$), and 
 $\Omega_\mathrm{lso}$ is given by $\Omega_{\rm lso} = j_{\rm lso} / X^2$,
 where $j_{\rm lso}$ is the specific angular momentum in the last stable orbit of
 the Schwarzshild BH (e.g., \citet{bardeen,prog05,lopez09}).
 By changing $\alpha$ in the range 
of $0.6 \le \alpha \le 1.2$, we compute 7 non-magnetized models of 
$\alpha$ = 0.6, 0.7, 0.8, 0.9, 1.0, 1.1, and 1.2, 
by which they are labeled as models J0.6, J0.7, and so on (see Figure 1). 
 To see the effects on magnetic fields, we compute two more models, in which 
 the initial magnetic field ($B_0 = 10^{10}$ G or $10^{11}$ G) are added
 to model J0.8 (model J0.8B10 or model J0.8B11), 
 in which the initial field is assumed 
to be purely poloidal and also assumed to be uniform and 
parallel to the spin axis.

Figure 1 shows profiles of the initial angular momentum for some models.
As shown, our models are taken to possess much more rapid rotation 
especially in the range of 1 to 4 $M_{\odot}$ in the mass coordinate compared 
 to the original profile of model 35OC (black dashed line). We set such a rotational
 profile otherwise the accretion disk cannot survive later than $\sim$2 s after the 
onset of gravitational collapse in our simulation. Later on, 
the accretion disk is swallowed to the 
central object due to the neutrino cooling which deprives the pressure support
 in the disk. In this case, one cannot account for 
 the duration of long bursts, the activity of which is typically longer than 
 $\sim 2$ s, and can last up to tens of minutes. So we experimentally choose
 to adjust the initial angular momentum as in Equation (1). 


 Although our models do have higher initial 
 angular momentum, the deviation may not be so serious (compare
 the light blue line that is for the angular momentum amplified 
 by a factor of 2 in the original progenitor) considering uncertainties
 in stellar evolution calculations (such as in the treatment of mass-loss,
 weak interactions, and fluid instabilities (e.g., convection, semiconvection, 
 rotation, and magnetic fields)). 
To mimic the original progenitor structure, we take $X_0$ to be the size of the 
Fe core ($\approx 3000 \km$).
 By setting $\alpha = 0.8$, $\Omega_0 \approx 1 \radps$, the above profile becomes 
 most close to the original profile for the mass coordinates larger
 than $5 M_{\odot}$. As a side-remark, a most prevailing way is to tune 
 the initial angular momentum to satisfy its local centrifugal 
 force to be several percent-levels of the local gravitational binding energy
 (e.g., \cite{macf99}). Our modeling may be more realistic in the sense that 
  the assumed initial angular momentum profiles capture the basic trend 
obtained in the current progenitor models.

\subsubsection{Hydrodynamics}
As already mentioned in the introduction,
a number of Newtonian and SR 
collapsar simulations in the phase 2 have conventionally focused on
 the evolution of the collapsar disks by treating the central BH as an absorbing 
boundary.
 Following the tradition of setting the boundary from the beginning of the 
 simulation,  we also initially impose an absorbing inner-boundary condition
 at the radius of $max (10 \km ,2r_g)$ 
with $r_g$ is the Schwarzschild radius 
that is estimated by the accumulating mass inside the inner-boundary. Then 
 we solve the dynamics outside the inner boundary up to the outer boundary 
 of the computational domain (30000 km in radius) by our 
 SRMHD code assuming axisymmetry and equatorial symmetry 
(see \citet{hari09} for more details).
 In our 2D simulations, spherical coordinates are employed
 with logarithmic zoning in the radial direction ($r$) 
and regular zoning in the polar direction ($0\le\theta\le\pi/2$).
 The computational domain is covered by 300($r$) $\times$ 40($\theta$) mesh points. 
 Regarding the microphysics, a realistic nuclear equation of state 
 by \citet{shen98} is included and the neutrino cooling is treated by a multi-flavor 
leakage scheme (e.g., \citet{epst81,ross03,kota03a}, see \citet{taki09} for more 
 details). The gravitational potential is estimated by the sum of the 
Paczynski \& Witta-type potential which mimics the gravitational pull 
from the central BH and 
 the self-gravity of material outside the excised region that is determined 
by the Poisson equation (see equations (5) and (6) in \citet{hari09}).
In this paper, we examine 
numerical models without viscosity like the 
alpha prescription (see, however, \citet{macf99,lindner,camara}).

\subsection{Extraction of Gravitational Waveforms}\label{2.2}

\subsubsection{GWs from matter motions}

To extract the gravitational waveforms from matter motions and magnetic fields,
 we employ the stress formulae derived in \citet{taki_kota,kota04a}. For convenience,
 we shortly summarize them in the following.

In our axisymmetric case, the non-vanishing quadrupole term
  is only the plus mode ($h_{+}$) in the metric perturbation (e.g.,
 \citet{mm}), which
 can be written as, 
\begin{equation}
h ({\mbox {\boldmath  $X$}},t) 
 = \frac{1}{8}\sqrt{\frac{15}{\pi}}\sin^2 \theta \frac{A^{E2}_{20}\left(t - 
\frac{R}{c}\right)}{R},
\label{htt}
\end{equation}
(e.g., \cite{thorne}).
The quadrupole amplitude of matter GWs;
${A_{20}^{\rm{E} 2}}_{\rm matter}$ that consists of the following three 
parts,
\begin{equation}
  {A_{20}^{\rm{E} 2}}_{\rm (matter)}  =  {A_{20}^{\rm{E} 2}}_{\rm (hyd)} + 
  {A_{20}^{\rm{E} 2}}_{\rm (grav)} + {A_{20}^{\rm{E} 2}}_{\rm (mag)}.
\label{A20}
\end{equation}
 The first term in Equation (\ref{A20}) represents the contribution from non-spherical hydrodynamic motions,
 which is expressed by
\begin{eqnarray}
 {A_{20}^{\rm{E} 2}}_{\rm(hyd)} &=& \frac{G}{c^4} \frac{32 \pi^{3/2}}{\sqrt{15 }} 
\int_{0}^{1}d\mu 
\int_{0}^{\infty}  r^2  \,dr \nonumber \\ 
& & \times \rho_{*}W^2( {v_r}^2 ( 3 \mu^2 -1) + {v_{\theta}}^2 ( 2 - 3 \mu^2)
 - {v_{\phi}}^{2} - 6 v_{r} v_{\theta} \,\mu \sqrt{1-\mu^2}),
\label{quad}
\end{eqnarray}
 in which  $\rho_{*}$ is the effective density defined as, 
\begin{equation}
\rho_{*}= \rho + \frac{e + p + |b|^2}{c^2}.
\end{equation}
 Here $\rho$, $e$, $p$, $c$, and $G$, denotes the baryon density, internal energy, 
pressure, the speed of light, and gravitational constant, respectively, 
and $|b|^2 = b^{\mu} b_{\mu}$ 
 is related to the energy density of the magnetic fields 
with $b_\mu$ representing the magnetic field 
in the laboratory frame (e.g., \citet{taki09}). 
 $W= 1/\sqrt{1-v^kv_k}$ is the Lorentz boost factor with $v_k$ denoting the
 spatial velocity in the spherical coordinates ($i~= r, \theta,\phi)$. 
$\mu = \cos \theta$ is a directional cosine.
The second term in Equation (\ref{A20}) represents the contribution from the 
 gravity
 as,
\begin{eqnarray}
 {A_{20}^{\rm{E} 2}}_{\rm(grav)} &=& \frac{G}{c^4} \frac{32 \pi^{3/2}}{\sqrt{15 }} 
\int_{0}^{1}d\mu  \int_{0}^{\infty}  r^2  \,dr\nonumber \\ 
& &
 \times \left[\rho h (W^2+(v_k/c)^2) + 
 \frac{2}{c^2}\left( p+\frac{\left|b\right|^2}{2}\right)
-\frac{1}{c^2}\left((b^{0})^2+(b_{k})^2 \right)\right]\nonumber \\
& & \times \left[- r \partial_{r} \Phi (3 \mu^2 -1) + 3 \partial_{\theta} \Phi \,\mu
\sqrt{1-\mu^2}\right],
\label{grav}
\end{eqnarray}
 where $\Phi$ denotes the gravitational potential of the self-gravity.
The last term in Equation (\ref{A20}) is the contribution from the magnetic fields as,
\begin{eqnarray}
 {A_{20}^{\rm{E} 2}}_{\rm(mag)} &=& - \frac{G}{c^4} \frac{32 \pi^{3/2}}{\sqrt{15 }} 
\int_{0}^{1}d\mu  \int_{0}^{\infty}  r^2  \,dr\nonumber \\ 
& &
\times [{b_r}^2 ( 3 \mu^2 -1) + {b_{\theta}}^2 ( 2 - 3 \mu^2)
 - {b_{\phi}}^{2} - 6  b_{r} b_{\theta} \mu \sqrt{1-\mu^2}].
\label{mag}
\end{eqnarray}

Here, we write the total gravitational amplitude as follows for later convenience,
\begin{equation}
h^{\rm TT}_{\rm (matter)} = h^{\rm TT}_{\rm(hyd)} + h^{\rm TT}_{\rm(mag)} + h^{TT}_{\rm(grav)}, 
\label{total}
\end{equation}
where the quantities of the right hand of the equation are defined by combining
Equations (\ref{htt}) and (\ref{A20}) with Equations (\ref{quad}), (\ref{grav}), and (\ref{mag}). 
  Note that by dropping $O(v/c)$ terms, 
the above formulae reduce to the conventional quadrupole formula employed in the Newtonian simulations (e.g., \cite{mm}). In the collapsar disk that we pay attention 
 in this work, the condition of $v_{\phi} \gg v_{\theta}, v_{r}$ 
in Equation (\ref{quad}) is generally satisfied inside the disk. 
If the disk is perfectly in a 
 stationary state, which means the centrifugal force ($\sim \rho v_{\phi}^2$)
 balance with the gravitational forces (e.g., Equation (\ref{grav})), no GWs
 can be emitted. As will be explained later, the disk attains mass continuously 
due to mass-accretion whose specific angular momentum increases outward (e.g., Figure 1).
 This is the primary reason of generating non-zero matter GWs from 
axisymmetrically but dynamically rotating collapsar disks.
 In the following computations, we assume that the observer is located in 
the equatorial plane ($\theta = \pi/2$ in Equation (\ref{htt})), and also that
 the distance to the GW source is comparable to 
 nearly GRB-associated core-collapse supernovae ($R = 100 ~\rm{Mpc}$) 
unless stated otherwise.

\subsubsection{GWs from anisotropic neutrino emission}
To compute the gravitational waveforms from anisotropic neutrino 
radiation, we follow the formalism pioneeringly proposed by \cite{epstein,muyan97}.
In the case of our 2D axisymmetric case, 
 the only non-vanishing component is the plus mode for the equatorial observer, 
\begin{eqnarray}
h_{\nu} &=&   \frac{4G}{c^4 R}  \int_{0}^{t} dt^{'}
\int_{0}^{\pi}~d\theta'~\Phi(\theta')~\frac{dl_{\nu}(\theta',t')}
{d\Omega'},
\label{tt}
\end{eqnarray}
 where $\Phi(\theta^{'})$ depends on the angle measured from the symmetry axis 
($\theta^{'}$) 
\begin{equation}
\Phi({\theta^{'}})=  \pi \sin \theta^{'} ( - 1 + 2 | \cos  \theta^{'}| ).
\label{graph1}
\end{equation} 
As given in Figure 1 of \citet{kotake07}, this function
 has positive values in the north polar cap for $0 \leq \theta' \leq 60^{\circ}$ and in 
the south polar cap for $120^{\circ} \leq \theta' \leq 180^{\circ}$, 
but becomes negative values between $60^{\circ} < \theta' < 120^{\circ}$. 
To determine the anisotropy in neutrino emission (i.e., $dl_{\nu}/d\Omega$
 in Equation (\ref{tt})), we perform a ray-tracing analysis, which was 
 first proposed to be applicable 
in the Newtonian gravity 
 \citep{kotake09,kotake_ray,kotake11} and later improved to be treatable in SR and GR 
\citep{harisr,hari_grrt}.

Applying the formalism in \citet{lindq66}, 
the Boltzmann equation for the neutrino occupation probability 
$f_{\nu}(\epsilon_{\nu},\Omega)$ for a given neutrino energy ($\epsilon_{\nu}$) 
along a specified direction of $\Omega$ can be 
 expressed as,
\begin{equation}
 \frac{df_{\nu}(\epsilon_{\nu},\Omega)}{d\lambda} = n [Q_e(1-f_{\nu}) - \kappa f_{\nu}] = n[Q_e - \kappa^{*} 
f_{\nu}],
\label{render_neutrino}
\end{equation}
 where $n(\mbox{\boldmath $x$})$ is the proper number density of the external 
 medium with which neutrinos interact and thus measured in its own local rest 
frame, $\epsilon_{\nu}$ is the neutrino energy 
measured in the local proper frame,
$\lambda$ denotes an affine parameter along the geodesics, 
 $Q_e$ and $\kappa$ represents neutrino emissivity and absorptivity, and
 $(1-f)$ represents the Pauli blocking term (e.g., \citet{hari_grrt} for more
 detailed information to derive the equation). 
Here we consider only $\nu_e$ and $\bar{\nu}_e$ for simplicity, and
 the energy and momentum transfer via neutrino scattering are neglected, which is not only difficult to be treated by the ray-tracing 
technique but also a major undertaking in the radiative transport problem in general.
In the final expression of Equation (\ref{render_neutrino}), $\kappa^{*}$ is defined to represent the effective absorptivity.
As for the opacity sources of neutrinos,
electron capture on proton and nuclei, 
positron capture on neutron, neutrino scattering with nucleon and nuclei,
 are included \citep{full85,taka78,bruenn}. Here $\kappa^{*}$ is
 estimated as $\kappa^{*} = \Sigma[n_{\rm target}\cdot \sigma(\epsilon_{\nu})]$ with 
 $n_{\rm target}$, $\sigma(\epsilon_{\nu})$ being the target number density of each 
reaction and the corresponding cross sections, respectively.

According to \citet{zink08}, the formal solution of Equation (\ref{render_neutrino}) can be given as
\begin{equation}\label{eq:rt3}
	f_{\nu}(\epsilon, \Omega) = \int^{\lambda_{\rm out}}_{\lambda_{0}} n(\lambda'') 
	Q_e(\lambda'', \epsilon_{\nu})  \exp\Biggl[{-\int^{\lambda_{\rm out}}_{\lambda''} n(\lambda') 
	\kappa^{*}(\lambda') d\lambda' }\Biggr] d\lambda'', 
\end{equation}
 which is referred to as the rendering equation of the radiation transport 
problem. We perform a line integral along the geodesics 
 from every point on the surface of neutrinospheres ($\lambda_{0}$) from 
 which neutrinos can escape freely, up to the outer-most boundary of 
 the computational domain ($\lambda_{\rm out}$).
 On the neutrinospheres, the neutrino distribution function is assumed to take a 
 Fermi-Dirac type  $f^{\rm FD} [= 1/({\epsilon_{\nu}}/{k_{\rm B} T}+1)]$ with a vanishing chemical potential\footnote{$k_{\rm B}$ is the Boltzmann constant}. For the neutrino energy bins ($\epsilon_{\nu}$), we use 16 logarithmically spaced 
energy bins reaching from 3 to 300 MeV.

The path integration in Equation (\ref{eq:rt3}) 
is done explicitly along the geodesics.
In doing so, we determine each integration step by restricting the maximum change 
of neutrino opacity for all the neutrino energy-bins to be less than 10 \%. 
 By using an adaptive-mesh-refinement approach also, our ray-trace code was 
 proved to safely pass several test problems (see \citet{hari_grrt} for more details).
 For the ray-tracing calculation, 300($r$) $\times$ 40($\theta$) $\times$ 32($\phi$) 
meshes points are cast. The radiation field typically changes more slowly than the 
hydrodynamics. We perform the ray-tracing calculation every 50 hydrodynamic step
  to reduce the computational cost.  Note that the obtained waveforms
  did not change significantly when the interval is varied as every 100 or 200 step.

 With $f(\epsilon_{\nu},\Omega)$ at the outer-most boundary, which is obtained by the 
above procedure, the emergent neutrino energy fluxes 
along the specified direction of $\Omega$ can be estimated,
 \begin{equation}
\frac{dl_{\nu}(\Omega,\epsilon_{\nu})}{d\Omega~dS}  = \int f(\epsilon_{\nu},\Omega)\cdot(c\epsilon_{\nu}) 
\cdot \frac{\epsilon_{\nu}^2 d\epsilon_{\nu}}{(2 \pi \hbar c)^3}.
\label{flux}
\end{equation}
 By summing up the energy fluxes with the weight of the area 
in the plane perpendicular to the rays (:$dS$), 
we can find $dl_{\nu}/d\Omega$ along the specified direction ${\Omega}$,
\begin{equation}
\frac{dl_{\nu}({\Omega})}{d\Omega} = \int \frac{dl_{\nu}({\Omega})}{d\Omega dS}~dS  
\label{final}
\end{equation}
Repeating the above procedures, $dl_{\nu}({\Omega})/d\Omega$ 
 can be estimated for all the directions, by which we can find the amplitudes of 
the GWs from neutrinos through Equation (\ref{tt}).

\section{Results}\label{results}

\begin{figure}[htbp]
\begin{center}
\epsscale{0.5}
\plotone{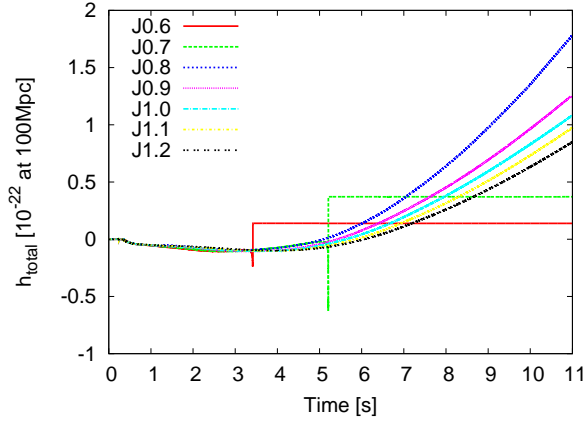}
\caption{Gravitational waveforms from the sum of neutrinos and matter motions 
for all the non-magnetized models. 
Note that a collapsar is assumed to be located at the distance of 100 Mpc.}
\label{f2}
\end{center}
\end{figure} 

\begin{deluxetable}{ccccc}
\tabletypesize{\scriptsize}
\tablecaption{Model Summary \label{table1}}
\tablewidth{0pt}
\tablehead{
\colhead{Model}
 & $\Delta t ({\rm s})$
& $|h_{\rm max}| (10^{-22})$
& $|h_{\nu}| (10^{-22})$
& $E^{\rm total}_{\rm GW} (\Ms c^2)$
}
\startdata
J0.6    & 3.41  & $0.47 $ & $0.14$   & $2.09 \times 10^{-6}$  \\
J0.7    & 5.22  & $1.15 $ & $0.37$   & $5.12 \times 10^{-5}$  \\
J0.8    & 11.0  & $2.47 $ & $1.80$   & $3.91 \times 10^{-3}$  \\
J0.9    & 11.0  & $3.25 $ & $1.22$   & $5.72 \times 10^{-3}$  \\
J1.0    & 11.0  & $2.86 $ & $1.07$   & $4.45 \times 10^{-3}$  \\
J1.1    & 11.0  & $2.54 $ & $0.98$   & $3.56 \times 10^{-3}$  \\
J1.2    & 11.0  & $2.23 $ & $0.86$   & $3.26 \times 10^{-3}$  \\
J0.8B10 & 2.97  & $0.18 $ & $0.037$  & $1.73 \times 10^{-5}$  \\
J0.8B11 & 0.53  & $0.028$ & $0.0065$ & $9.00 \times 10^{-8}$  \\
\enddata
\tablecomments{
$\Delta t$ represents the simulation time.
$|h_{\rm max}|$ represents the total GW amplitudes at maximum during the
simulation
time, while $h_{\nu}$ is the neutrino-originated GW at the end
of the simulations.
$E^{\rm total}_{{\rm GW,}}$ is the total radiated energy in the form of the
 GWs in unit of $M_{\odot} c^2$. The distance to the source is assumed to
be 100 Mpc. }
\end{deluxetable}
 
\begin{figure}[htbp]
\begin{center}
\epsscale{.97}
\plottwo{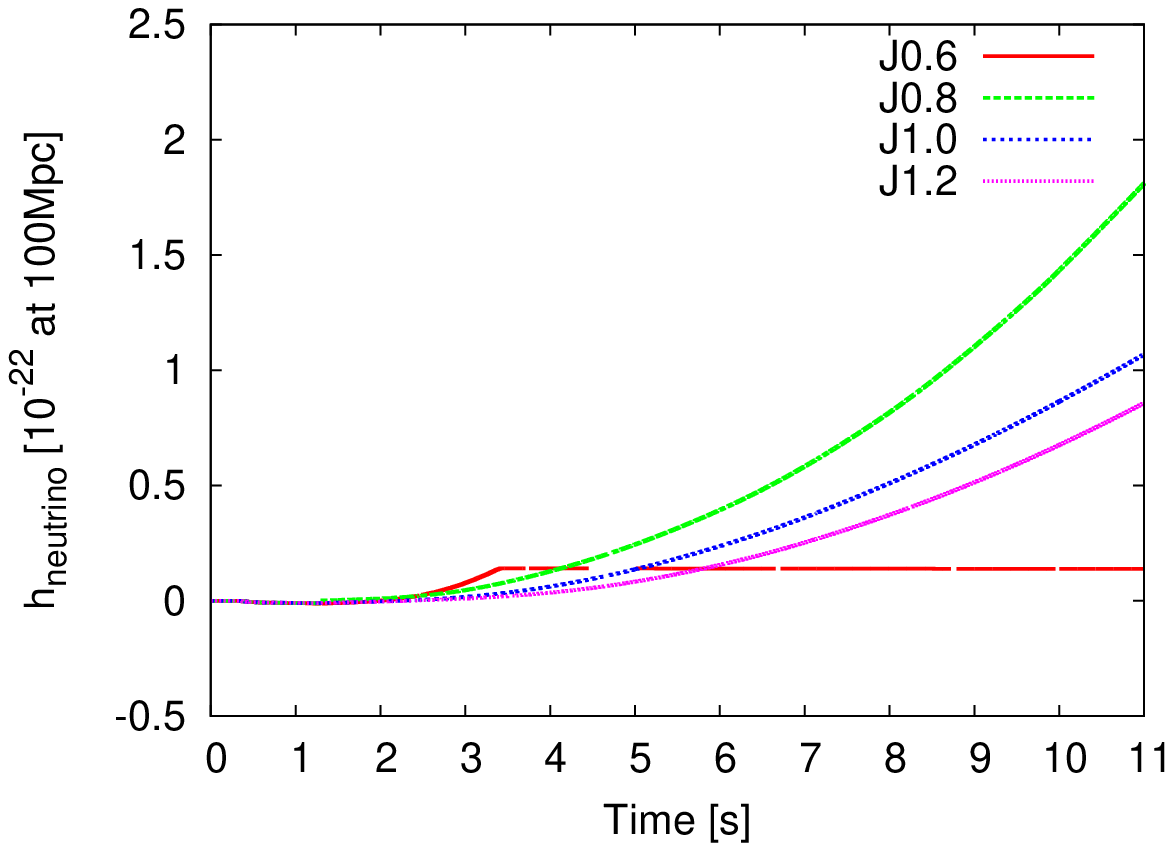}{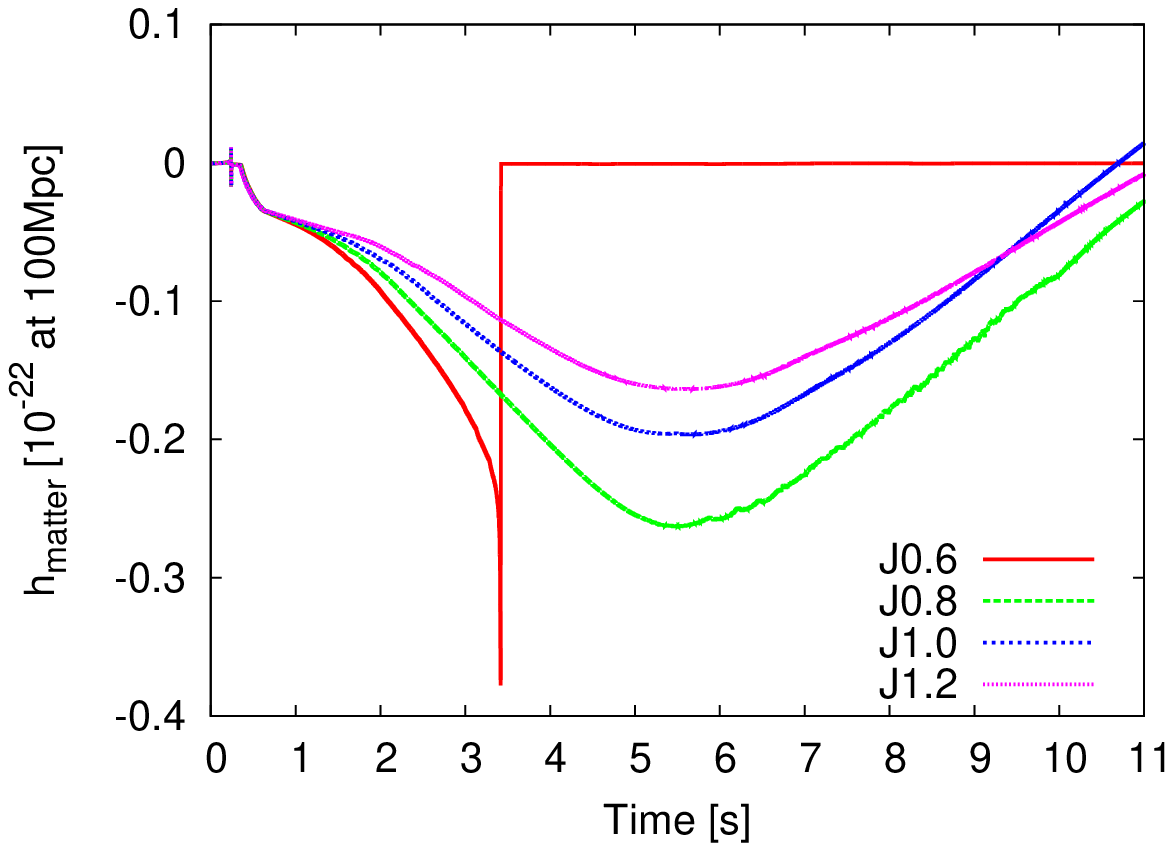}
\caption{Similar to Figure \ref{f2} but for the gravitational waveforms
  only from anisotropic neutrino emission (left panel, e.g.,
  Equation (\ref{tt})) and only from matter motions 
(right panel, e.g., Equation (\ref{total})).
 Note 
 that only the selected models are drawn in this figure not to make it
 messy.}
\label{f3}
\end{center}
\end{figure} 

First we pay attention to the properties of GWs as well as the hydrodynamic features in
  models without magnetic fields.

 Figure \ref{f2} shows the total GW amplitudes for all the computed
 models without magnetic fields (see also table 1 for a model summary). As seen, 
the total GW amplitudes generally
 show a positively growing feature with time. In addition,
a sudden disappearance in the signals can be seen for some slowly rotating models 
 (models J0.6 (red line) and J0.7 (green line)). Figure \ref{f3} shows 
 the gravitational waveforms contributed only from anisotropic neutrino emission
 (left panel) and only from matter motions (right panel), respectively. 
 The positively growing trend is shown to come from the neutrino contribution 
(left), which is much larger than the matter contribution (right).
 To understand these properties, we first briefly summarize a hydrodynamic evolution of 
 our 2D non-magnetized models in section \ref{3.1}.
And then in section \ref{3.2}, we move on to analyze the reason of the 
 positive growth in the neutrino GWs by performing the ray-tracing analysis. 
After that, we analyze the matter GWs paying particular attention to the 
 magnetic effects in section \ref{3.3}. We then discuss their detectability 
in section \ref{3.4}.


\subsection{Hydrodynamic Features}\label{3.1}

\begin{figure}[hbtp]
\epsscale{0.5}
\plotone{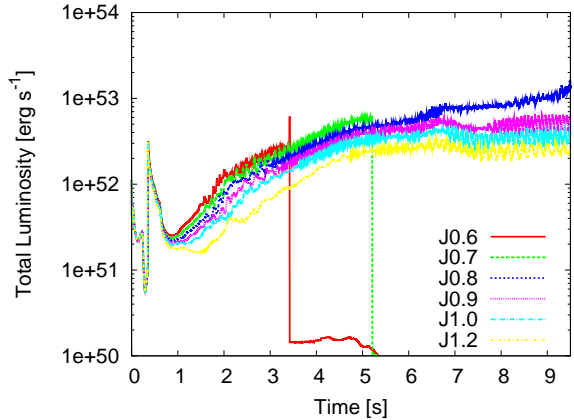}
\caption{Time evolution of neutrino luminosities (the sum of all the neutrino species) 
for models J0.6, J0.75, J0.8,
  J0.85, J1.0, and, J1.2, respectively. Note that the time is measured from 
the epoch when the simulations are started (i.e., the onset of gravitational 
 collapse). }
\label{f4}
\end{figure} 

To capture hydrodynamic features in our models, Figure \ref{f4} shows
 the evolution of neutrino luminosities for some selected models.
 In $t \sim 1.0$ s after we start our simulations ($t \equiv 0$ s), all 
the models experience rapid infall and the subsequent shock formation 
in the center, which leads to a drastic increase 
and the subsequent decrease in the neutrino luminosities\footnote{Note that the shock formation is not because the central density 
 exceeds the nuclear density, but because
 the matter pressure becomes so high due to compression that it pushes back the 
ram pressure of accreting material in the vicinity of the inner boundary. In contrast
 to the state-of-the-art simulations of BH-forming
 core-collapse supernovae (e.g., \citet{sumiyoshi,tobias,ott2011}),
 the inability of capturing dynamics correctly especially before the BH formation, is 
 one of major drawbacks in collapsar simulations in general.}. 

\begin{figure}[htbp]
\begin{center}
\includegraphics[scale=0.4]{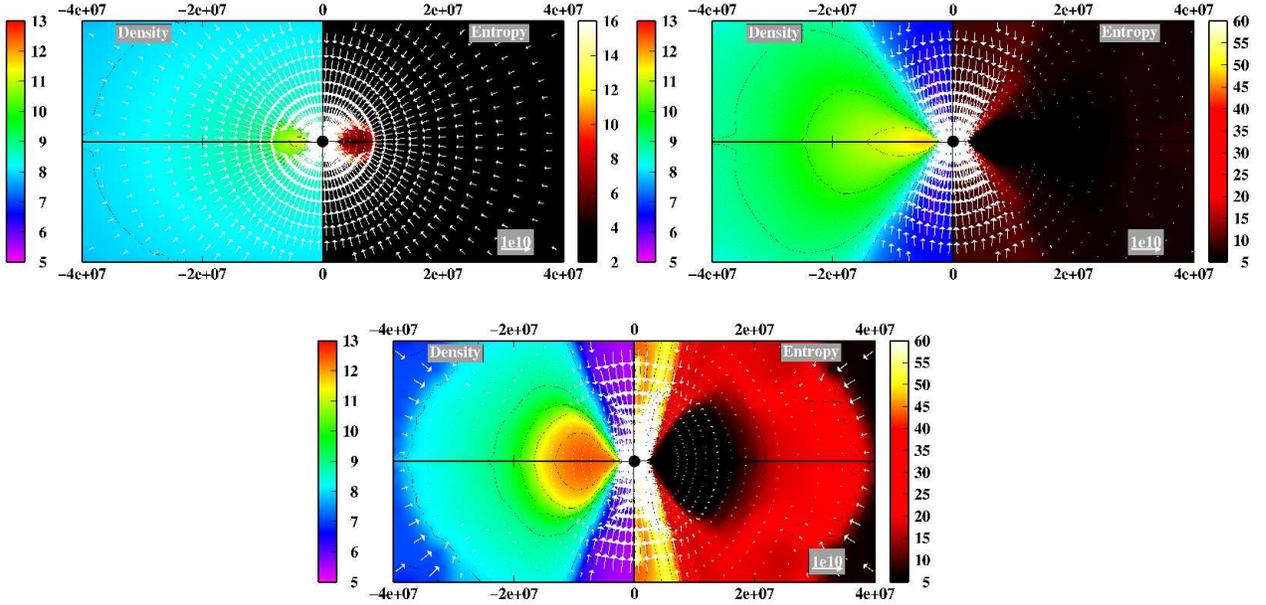}
\caption{Three snapshots characterising the hydrodynamic evolution for model J0.8.
 The pair panels (top left) are for $t=0.42$ s, and the top right, bottom panels
 are for $t=1.89$ and $t=9.00$ s, respectively. In the pair panels, the left half 
 in the top panel shows the logarithmic contour of 
 density (in$\gpcmc$), while the right half is for entropy per nucleon in unit of 
 the Boltzmann constant($k_{\rm B}$).
 The velocity fields are drawn by the white arrows, and 
the length is normalized by the scale shown in top right edge of the box 
(i.e. $10^{10} {\rm cm}~{\rm s}^{-1}$). 
 The central black circle represents the inner boundary 
of our computations.}
\label{f5}
\end{center}
\end{figure} 

Here we take model J0.8 as a reference, because the precollapse angular momentum is 
 adjusted to be closest to the original progenitor.
 The top left panel in Figure \ref{f5} show a snapshot at $t = 0.42$ s 
near the shock formation. 
Note that the accretion mass in this epoch is typically greater 
than 2$\sim 3 \Ms$ in the center. The maximum mass of the neutron star of the 
 Shen equation of state is in the same mass range (\citet{shen98}, 
see also \citet{oconnor,kiuchi08}). And recent full GR simulations show that the 
 mass of the central object just before collapsing to a BH is typically 
$\lesssim 2.3 M_{\odot}$ (e.g., \citet{ott2011}) with its typical radius of several km
 (as inferred from their Figure 3). These evidences might support our very crude 
assumption of the prompt BH formation that is modeled by setting the BH initially
 in the center, although such assumption can be only tested by full GR simulations
 using the same progenitor model.
 Later on, the density configuration (compare the left-half in
 each panel) deforms to be more oblate with time due to accretion of material
 with higher angular momentum outside (e.g., Figure \ref{f1}).
As will be explained in the later section, 
the luminous accretion disk is the primary source of the 
GWs from anisotropic neutrino emission.

The top right panel in Figure \ref{f5} shows a snapshot at $t=1.89$ s.
 The bluish regions near along the spin axis of the accretion disk (left-half, density)
 correspond to the so-called polar funnel regions. The entropy becomes highest 
 near the surface of the accretion disk due to the shock heating 
when the accreting material hits the wall of the disk (right panel).
Comparing Figure \ref{f3} (left panel) to Figure \ref{f4},
 the GWs from neutrinos deviate from zero 
typically later than $t \gtrsim 2-3 $ s, when the (total) neutrino luminosities become
 as high as $\sim 10^{52}$erg/s. Later on, the neutrino luminosities show a gradual 
increase with time, 
reflecting the increase in the mass accretion to the newly formed accretion disk. 

\begin{figure}[hbtp]
\begin{center}
\includegraphics[scale=0.4]{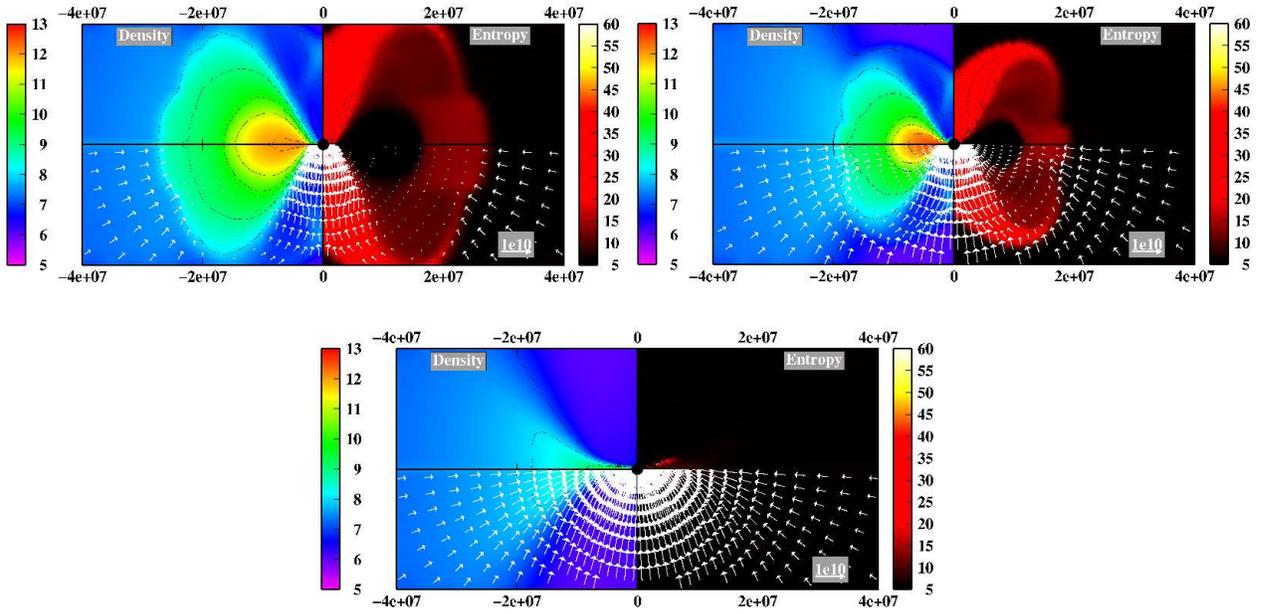}
\caption{Same as Figure \ref{f5} but for model J0.6.
 The top left, top right, and bottom panel is for $t=3.09,~3.42,~3.43$ s, 
 respectively. For this model, the accretion disk is 
 finally absorbed into the central BH due to its small initial angular momentum 
 (bottom panel), leading to a sudden decrease in the neutrino luminosity (e.g., 
 Figure \ref{f4}).}
\label{f6}
\end{center}
\end{figure} 

\begin{figure}
\begin{center}
\includegraphics[scale=0.8]{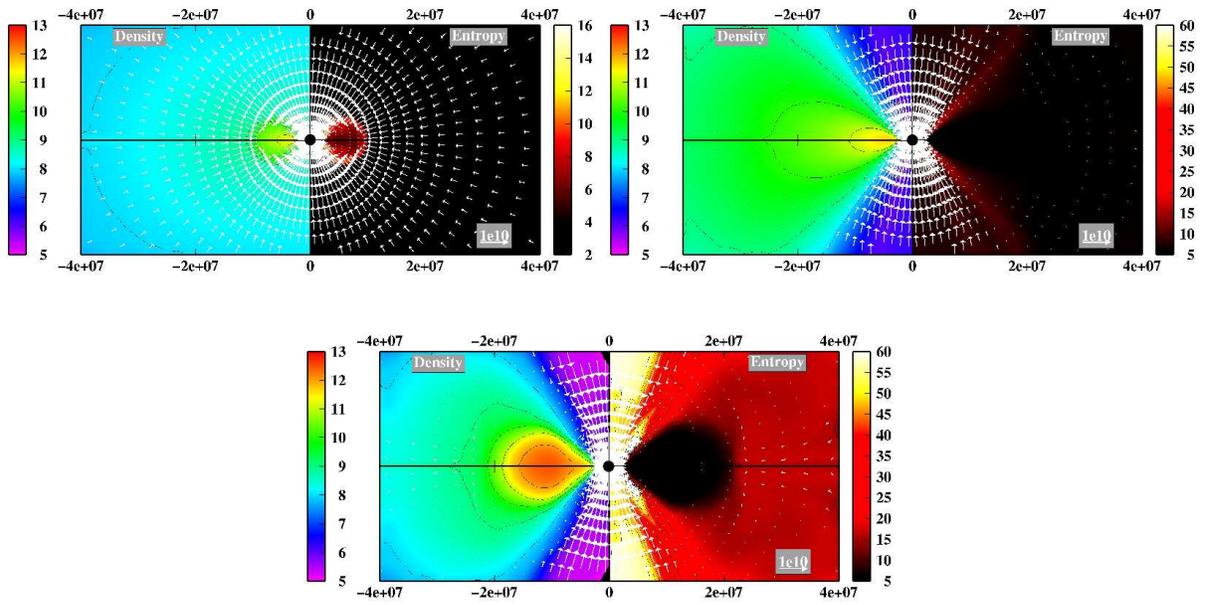}
\caption{Same as Figure \ref{f3} but for model J1.1.
 The top left, top right, and bottom panel is for $t=0.5,~2.4,~9.0$ s, 
 respectively. Due to a more larger initial angular momentum than that 
 for Figure \ref{f5}, the accretion disk is deformed to be more oblate.}
\label{f7}
\end{center}
\end{figure} 

In Figure \ref{f4}, the neutrino luminosity for models J0.6 (red line) 
and J0.7 (green line) is shown to steeply decrease at $\sim$ 3.3 s and 7.5 s, 
respectively.  This is because the accretion disk 
 is swallowed to the center (bottom panel in Figure \ref{f6}) 
 mainly because the pressure support in the accretion disk is reduced by 
  the neutrino cooling.
 The disappearance of the accretion disks is also the reason of the sudden decrease 
 in the GW signals observed both in the neutrino and matter sectors 
(e.g., Figures \ref{f2} and \ref{f3}).
 Before the disappearance, the accretion disk
 is observed to show a rapid expansion and contraction (as indicated by the 
 top panels in Figure \ref{f6}). At the same time, the mass flux on the 
Lagrange point around the disk changes violently, which may 
be related to the so-called runaway instability 
(e.g., \citet{abram83,font02}). 
Except for these slowing rotating models, the neutrino luminosities 
gradually settle to be nearly constant typically later than $t \sim 6$ s 
(Figure \ref{f4}).
 The saturation of the neutrino luminosity is because the 
 disk is already deleptonized and the neutrino emission there is suppressed
 (e.g., \citet{hari09} for more details).
 The neutrino luminosities at this epoch become as high as $10^{52 \sim 53} \ergps$,
 which touches the level of $\sim 10 \%$ of the 
accretion luminosities (see also \citet{chen,seki11}). 
Apparently the accretion disk is neutrino-cooling dominated. 

 From Figure \ref{f4}, the neutrino luminosity of model J0.8 is shown to be highest 
(e.g., blue line). Models with higher initial angular momentum have 
more extended disks due to larger centrifugal forces (compare the bottom
 panels in Figure \ref{f5} and \ref{f7}), leading
 to lower density and temperature in the disks. 
 This is the reason that the neutrino luminosities for models J1.0 (light blue
 line), J1.2 (yellow line in Figure \ref{f4}) are lower in this order.
 Reflecting this, the GWs from neutrinos 
 become highest for model J0.8 (left panel of Figure \ref{f3}, see also $|h_{\nu}|$
 in Table 1). Regarding the matter GWs, it should be noted that their maximum absolute 
amplitudes are obtained 
  not for the most rapidly rotating model (model J1.2), but for mode J0.8 (right panel
 of Figure \ref{f3}). This situation may be akin to the GW signals emitted 
 near core-bounce in core-collapse supernovae (see \citet{kota06,kotake11,ott_rev} for 
 recent reviews).
 Too much initial angular momentum works to suppress the matter 
compression leading to the smaller mass-quadrupole moment.
 As a result, the matter GWs becomes maximum for models with moderately rotating
 models (e.g., \citet{mm,yama95,zwer97,kotakegw,shibaseki,ott,ott_prl,dimm02,dimmelprl,dimm08,
simon1}).

\subsection{GWs from Anisotropic Neutrino Emission}\label{3.2}
In this section, we move on to look more into detail the properties of gravitational 
waveforms mentioned in the previous section. By taking model J0.8 as a reference,
 we first focus on the neutrino GWs.

\begin{figure}[hbpt]
\begin{center}
\epsscale{.80}
\plotone{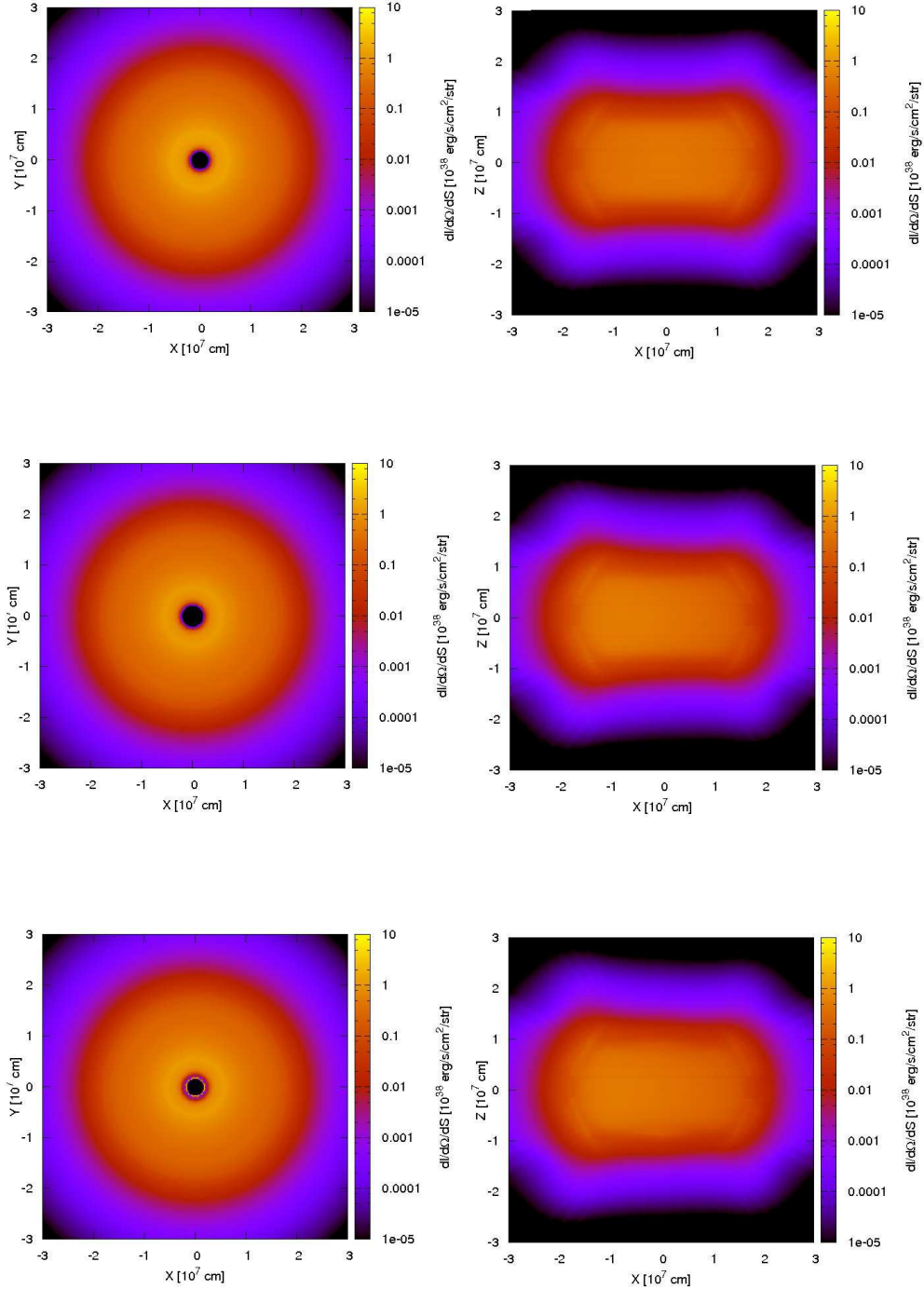}
\end{center}
\caption{The neutrino energy fluxes of
$dl_{\nu}/({d\Omega dS})$ (Equation(\ref{flux})) 
seen from polar (left) or equatorial direction
 (right panels), respectively for model J0.8 at $t=9.1$ s. The top and middle panels
 are calculated for the Minkowski geometry without or with special relativistic 
corrections, and the bottom panels are for the extreme Kerr geometry
 ($a=0.999$, see text for more details). Note that the $Z$ axis in the right panels conincides with the spin direction of the accretion disk.}
\label{f12}
\end{figure}

 Figure \ref{f12} shows the local neutrino energy fluxes 
(:$dl_{\nu}/({d\Omega dS})$, Equation (\ref{flux})) 
 at $t=9.0$ s (e.g., bottom panel of Figure \ref{f5}) 
seen from polar (left) or equatorial direction (right), respectively. Note 
 that the polar direction is taken to be parallel to the spin axis of the accretion disk.
 The top panel is for the Minkowski geometry, which corresponds to a 
 purely Newtonian case. Reflecting the shape of the accretion disk in axisymmetry,
 the contours of the neutrino flux are deformed to be oblate when seen from the equator
 (top right), while the neutrino flux seen from the polar direction (top left) 
 looks like a superposition of circles in a concentric fashion, 
the center of which corresponds to the spin axis. 
 Note that the central small cavity in the top left panel
 corresponds to the inner boundary of our SR simulation. 

\begin{figure}[hbpt]
\begin{center}
\epsscale{1}
\plottwo{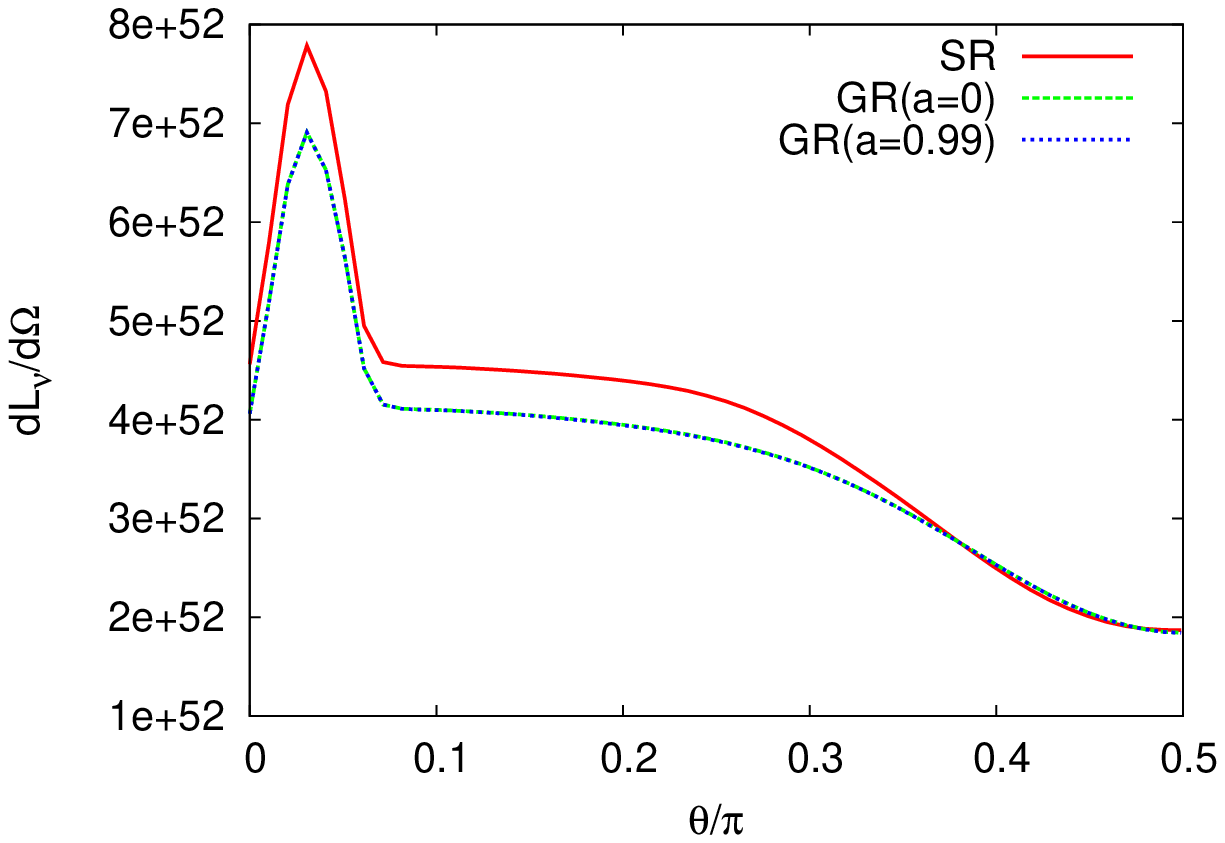}{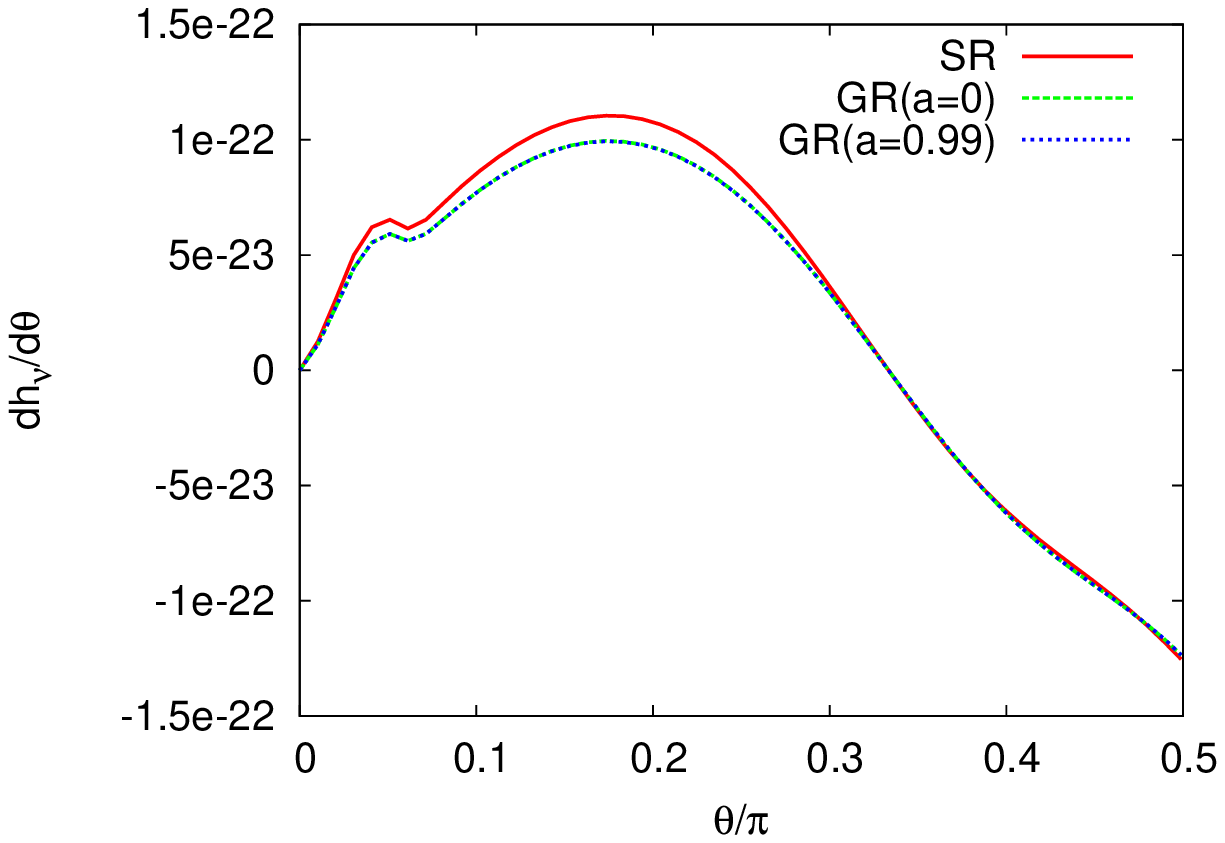}
\end{center}
\caption{The neutrino luminosity per angle (left panel,
 $dl_{\nu}/d\Omega$ (Equation (15)) and neutrino GWs per 
solid angle (right panel) for model J0.8 at $t=$9.0 s in the case of 
  the Minkowski (indicated by "SR"), Schwarzschild ("GR(a=0))", and 
extreme Kerr geometry ("GR(a=0.99))", respectively.
  }
\label{f13}
\end{figure}

 The middle panels are calculated for the Minkowski geometry including special 
relativistic corrections 
(see \citet{hari_grrt} for more detail).
Comparing the middle right to the top right panel, a clear difference is 
the left-right asymmetry in the middle right panel.
 This is due to the special relativistic beaming effects. 
Since the rotational velocity of the accretion disk 
is perpendicular to the polar direction, the special relativistic beaming 
effect suppresses the neutrino emission toward the polar region. As a result,
 the neutrino fluxes are made dark (right-hand side in the middle right panel) 
 because the direction of the rotating material is opposite to the direction to
 the observer. To study the GR effects on the neutrino luminosity,
 we place $4M_{\odot}$ in the central region, which mimics 
 the event horizon of the BH.  The effects of the BH spin are examined 
 by setting the Kerr parameter by hand as $a=0$ and $a=0.999$
for the Schwarzschild and the extreme Kerr geometry, respectively.
 The bottom panels of Figure \ref{f12} are for the extreme Kerr geometry,
 which however looks very similar to the middle panels.

For a more detailed comparison, the left panel of Figure \ref{f13} shows 
the neutrino luminosity per solid angle $dL_{\nu}/d\Omega$ for 
 the Minkowski (indicated by "SR"), Schwarzschild ("GR(a=0))", and 
extreme Kerr geometry ("GR(a=0.99))", respectively. 
 The most important message in this panel is that in every case, 
the neutrino luminosity seen from the direction near parallel 
 to the spin axis ($\theta = 0$) becomes higher 
than the one seen from the equatorial direction ($\theta = \pi/2$). This
 is because the cross section of the pan-cake like accretion disk seen from the spin 
axis becomes larger compared to the one seen from the horizontal direction 
(Figure \ref{f12}). 
 Remembering again that $\Phi(\theta^{'})$ 
in Equation (\ref{graph1}) is positive near 
the north and south polar caps,  the dominance of the polar neutrino luminosities
 make the neutrino GWs positive in the polar cap regions
 (right panel in Figure \ref{f13}). This is the reason that 
 the neutrino GWs increase monotonically with time (e.g., left panel of 
 Figure \ref{f3}). Here it
 is worth mentioning that from neutrino luminosity ($L_{\nu}$ in Figure \ref{f4}), 
their typical duration ($\Delta t_{\nu}$), anisotropy in the neutrino radiation 
($\alpha_{\nu}$, 
read from the left panel of Figure \ref{f13}), the GW amplitudes from neutrinos can 
 be estimated from Equation (\ref{tt}) as 
\begin{equation}
\label{gnu}
h_{\nu} \approx 10^{-22} \Bigl(\frac{\alpha_{\nu}}{0.2}\Bigr)\Bigl
(\frac{L_{\nu}}{10^{53}~{\rm erg}~{\rm s}^{-1}}\Bigr)
\Bigl(\frac{\Delta t_{\nu}}{10~{\rm s}}\Bigr)
\Bigl(\frac{D}{100 {\rm Mpc}}\Bigr)^{-1}, 
\end{equation}
 which is in good agreement with the numerical results (Figure \ref{f2}).

 Comparing to the SR with GR case in the left panel of Figure \ref{f13},
 the neutrino luminosity becomes 
smaller by $\sim 20-30 \%$ in the GR case due to 
 the GR redshift and also due to the bending effects.
 Comparing the green line with the blue line, the frame-dragging effect 
due to the BH spin barely affects the emergent neutrino luminosity. This is probably
  because the GR effect on the neutrino luminosity 
 becomes important only in the central regions very close to the BH 
(e.g., Figure 15 \citet{hari_grrt}). Our results indicate that 
 a ray-tracing calculation in the Schwarzschild geometry is 
 at least needed to accurately estimate 
 the neutrino GWs in the collapsar's environment.

\subsection{GWs from Matter Motions and Magnetic Fields }\label{3.3}
\begin{figure}[hbtp]
\begin{center}
\includegraphics[scale=0.8]{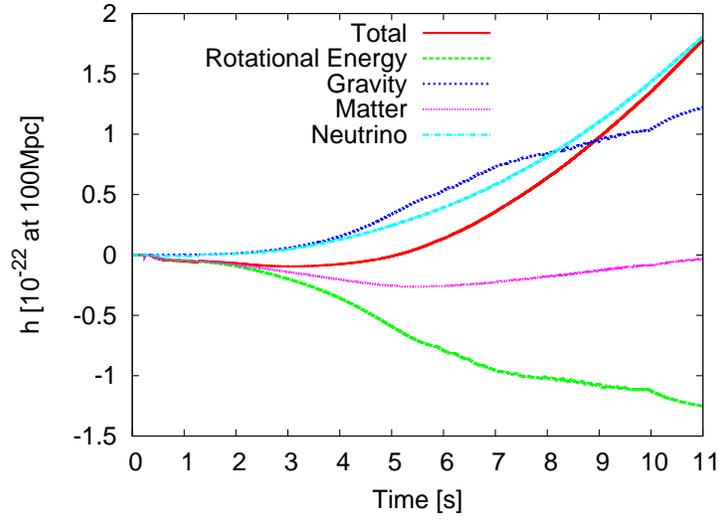}
\caption{The gravitational waveform for model J0.8.  "Total" 
denotes the total amplitudes (the sum of $h_{\rm (matter)}$ in Equation 
(\ref{total}) and $h_{\nu}$ in Equation (\ref{tt})),
 while "Matter", "Rotational Energy", "Gravity", and 
"Neutrino" represents the contribution from matter 
($h_{\rm (matter)}$ in Equation (\ref{total})), 
rotational energy (see text for the definition), gravity 
($h_{\rm (grav)}$ in Equation (\ref{total})), and neutrino emission 
($h_{\nu}$ in Equation (\ref{tt})), respectively.}
\label{f8}
\end{center}
\end{figure} 

Now we shortly analyze the matter GWs in model J0.8. As summarized in section \ref{2.2}, the matter GWs are estimated by 
the sum of the hydrodynamic and gravity parts in Equations (\ref{quad}, \ref{grav}) 
 for non-magnetized models.
 Among the kinetic energy terms ($\propto \rho v_i v_j$) 
in Equation (\ref{quad}), the largest contribution comes from the rotational energy 
(i.e., $ - \rho_{*}W^2v_{\phi}^2$), which is shown in Figure \ref{f8} (green line).
 In contrast to the negative contribution by this term,
 the gravity part (blue line labeled by "Gravity"
 in Figure \ref{f8}) is shown to make a positive contribution. 
 If a star rotates perfectly stationary, 
the centrifugal forces balance with the gravitational force, leading to no GWs.
In the collapsar disk studied in this work, the disk attains 
mass continuously due to mass-accretion with increasing its specific angular 
momentum outward. This is the reason why the disk is not perfectly 
stationary, leading to a non-zero GW emission from matter motions
(see pink line labeled by "Matter" in Figure \ref{f8}).
 However, their GW amplitudes
 are much smaller compared to those from anisotropic neutrino emission 
(compare the pink line with the light blue line (neutrino GWs) 
in Figure \ref{f8}). 

\begin{figure}[hbpt]
\begin{center}
\epsscale{1}
\plottwo{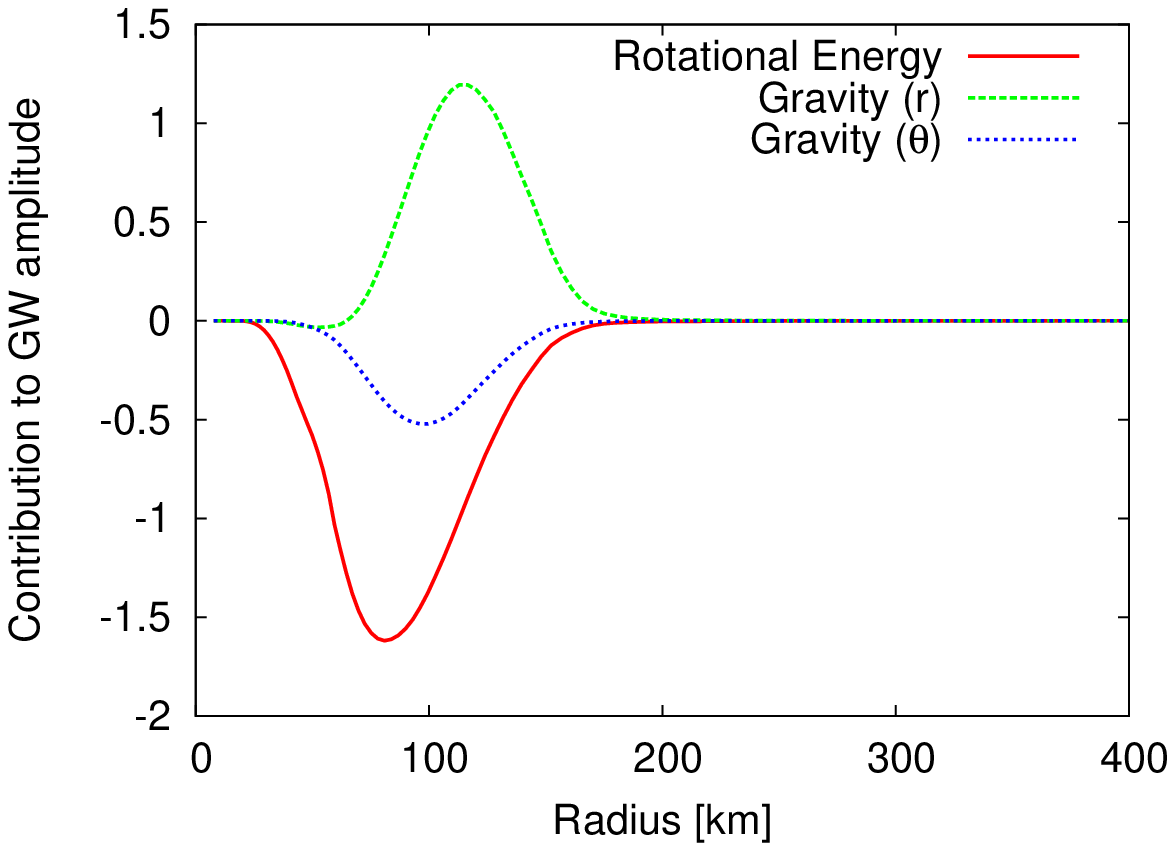}{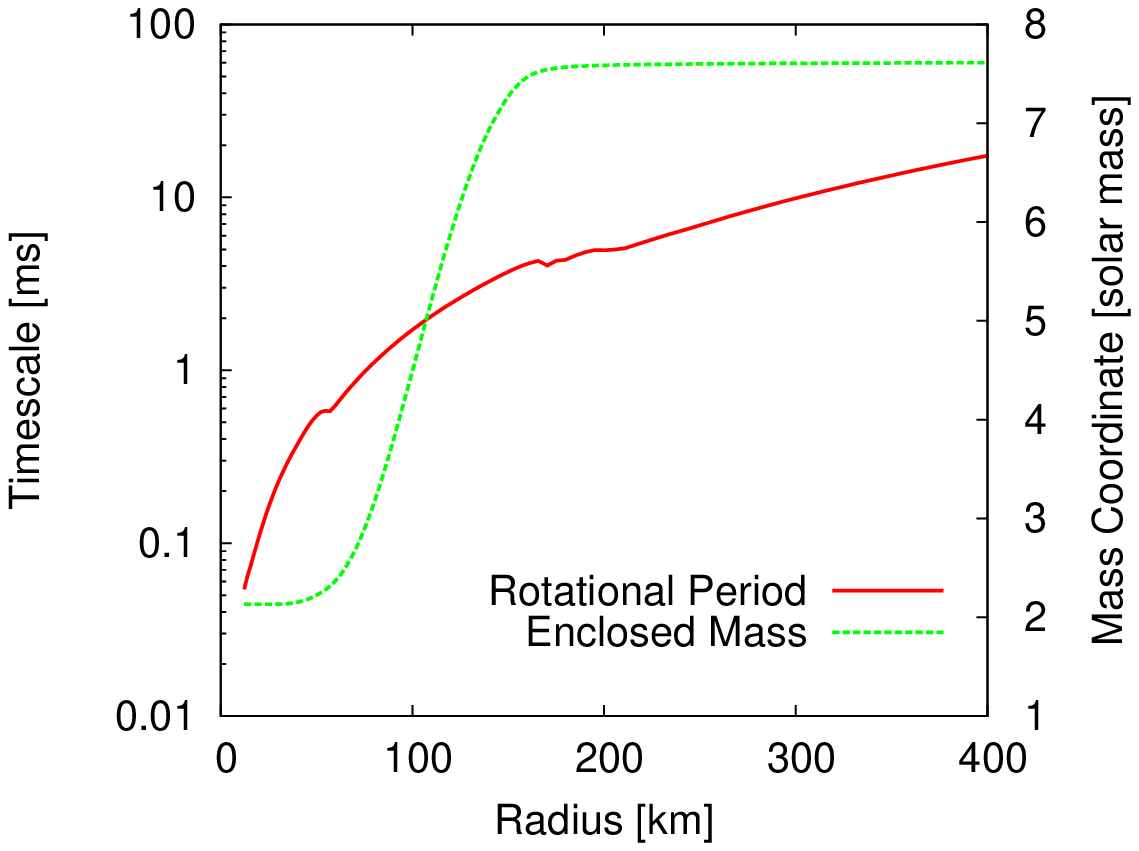}
\end{center}
\caption{Left panel shows a normalized contribution of each term in 
$ {A_{20}^{\rm{E} 2}}$ (e.g., Equation (\ref{A20}) as similar to Figure \ref{f8}) 
 as a function of radius for model J0.8 at $t=9$ s. 
"Gravity ($r$/$\theta$)" indicates 
 contributions from radial and lateral derivative of the gravitational potential
 (corresponding to the last two terms in Equation (6)).
 Right panel shows the radial 
profiles of the rotational period (red line) and the enclosed mass normalized by 
 $M_{\odot}$ (green line), respectively.}
\label{f9}
\end{figure}

The left panel of Figure \ref{f9} shows a normalized contribution of each term in 
${A_{20}^{\rm{E} 2}}$ at $t=9$ s for model J0.8, which 
is estimated by the volume integral of ${A_{20}^{\rm{E} 2}}$ within a given sphere
 enclosed by certain radius. The region in a radius between 50 km to 160 km
 is shown to contribute to produce GWs, which corresponds to a high density region
 in the accretion disk (see the bottom panel in Figure \ref{f5}). 
 It can be also shown that the radial gradient of the gravitational 
potential (green line indicated by "Gravity ($r$)" in the plot)
 closely cancels with the contribution from the centrifugal force (red line) 
in the disk.
 As shown in the right panel of Figure \ref{f9}, the enclosed mass there becomes 
as big as $\sim 4 M_{\odot}$ as a result of the hyperaccreting activity
 lasting $\sim$ 9 s till then,  and the typical rotational period there is the order 
of milliseconds (red line). 

 Putting these numbers to the standard GW stress formula (e.g., \citet{shap83}), 
 an upper bound of the matter GW amplitudes may be estimated as
\begin{eqnarray}
h_{\rm matter} &=& \frac{2G}{c^4 D} \ddot{I_{ij}} \sim \frac{2G}{c^4 D} \frac{M R^2}{T^2} 
\nonumber \\ &\lesssim & 
10^{-23} \Bigl(\frac{\epsilon}{0.1}\Bigr) \Bigl(\frac{100~{\rm Mpc}}{D}\Bigr)\Bigl(\frac{M}{4 M_{\odot}}\Bigr) \Bigl(\frac{R}{100~{\rm km}}\Bigr)^2 \Bigl(\frac{T}{4~{\rm ms}}\Bigr)^{-2},
\label{simple_estimate}
\end{eqnarray}
where $D$ is the distance to the source, $\ddot{I_{ij}}$ is the second
time derivative of the quadrupole moment of $I_{ij}$, $M$, $R$, and $T$ represents 
the typical mass and radius of the accretion disk and the
 timescale which we may take as the rotational period, respectively, and 
 $\epsilon$ is the degree of the nonsphericity, which we take
 optimistically as 10 $\%$ in the case of the accretion disk.
  This estimate, roughly consistent with the numerical results (e.g.,
 right panel of Figure 3), also shows that anisotropic neutrino emission,
 producing the GW amplitude on the order of $\approx 10^{-22}$ for a source of 100 Mpc,
  is the primary source in the long term evolution of collapsars ($\sim 10$ s) (see
 Equation (\ref{gnu})).

\begin{figure}[htbp]
\begin{center}
\includegraphics[scale=0.8]{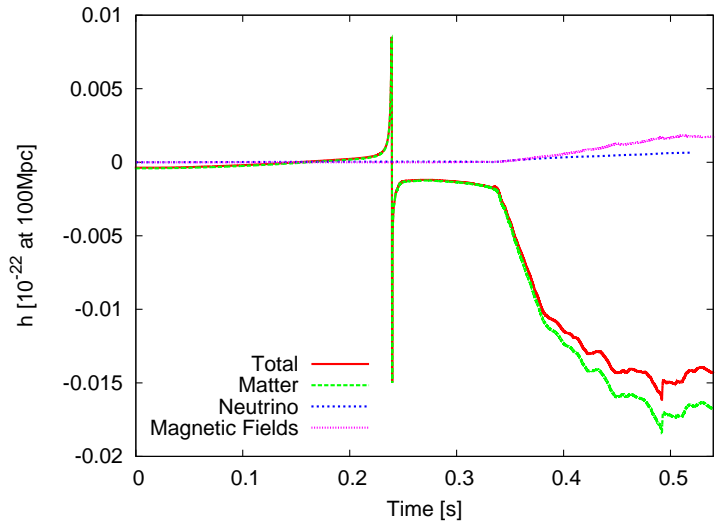}
\caption{Similar to Figure \ref{f2} but for model J0.8B11. The 
pink line represents the GWs from magnetic fields ($h_{\rm (mag)}$ in 
Equation (\ref{total})).}
\label{f10}
\end{center}
\end{figure}

 Now we discuss the effects of magnetic fields on the waveforms
  by taking model J0.8B11 that has a strongest initial magnetic field in our models. 
In Figure \ref{f10}, the pink line represents the GWs from 
magnetic fields (Equation (\ref{mag})). 
But, first of all, let us shortly comment on the burst-like feature of 
the green line 
(contributed from matter motions) at $t\sim 0.24$ s in the figure.
 This may look similar to the 
 burst of GWs emitted at the moment of the BH formation possibly followed by 
a damped sinusoidal oscillation (\citet{seidel90,seidel91}),
 but this can be only captured in full GR simulations
\citep{baiotti07,ott2011,seki11,kuroda12}. 
 As already seen in the luminosity plot of Figure \ref{f4}, the above burst simply 
corresponds to the shock formation at the center in our SR models.
Back to the main point, we analyze the increasing trend of the GWs from 
 magnetic fields in the following.

\begin{figure}[htbp]
\begin{center}
\includegraphics[scale=.7]{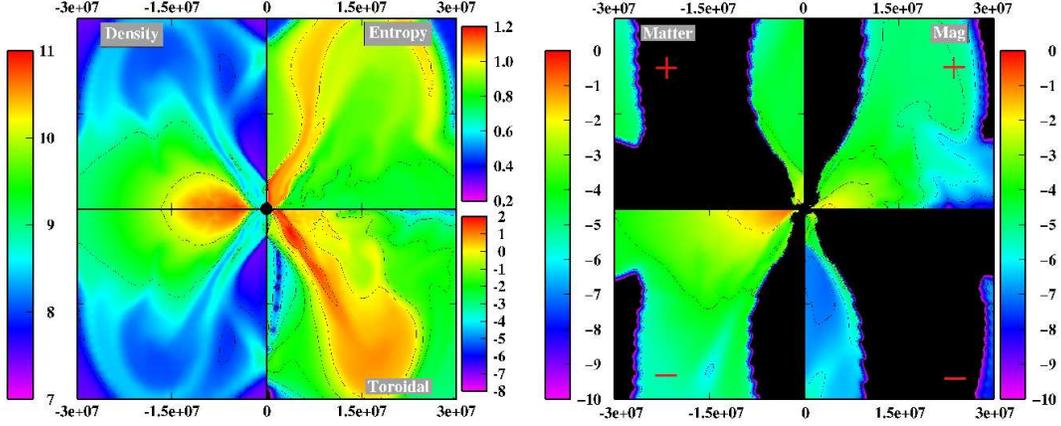}
\caption{Hydrodynamic configuration for model J0.8B11 at $t=$541 ms.
In the left panel, the logarithmic density (in$\gpcmc$, left-half),
 entropy($k_B$/baryon, right top), and the ratio of magnetic to the 
 matter pressure (right bottom) are shown. 
Right panel shows the sum of the hydrodynamic and gravitational parts 
(indicated by ``Matter'' in the left-hand side) and the magnetic part 
 (indicated by "Mag" in the right-hand side), respectively. The top and bottom 
panels represent the positive and negative contribution (indicated by (+) or (-))
 to ${A_{20}^{\rm{E} 2}}$,
 respectively (see text for more detail)}
\label{f11}
\end{center}
\end{figure}

The left panel of Figure \ref{f11} depicts a snapshot of density
  (left-half), entropy (right-top) and plasma $\beta$ (right-bottom, 
the ratio of magnetic to the matter pressure) at the final simulation time 
 ($t=541$ ms) for model J0.8B11.
 The high entropy regions in the slightly off-axis region (seen as reddish in the 
 top right panel) correspond
 to the magnetohydrodynamically-driven outflows that are pushed outwards
 by the twisted toroidal magnetic fields (also seen 
as reddish in the bottom right panel, indicated by "Toroidal").
  
The right panel of Figure \ref{f11} shows contributions
 to the total GW amplitudes (Equation (9)), in which the left-hand-side panels 
are for the sum of the hydrodynamic and gravitational part (indicated by ``Matter''),
 namely $\log\left(\pm\left[{A_{20}^{\rm{E} 2}}_{\rm (hyd)} + 
{A_{20}^{\rm{E} 2}}_{\rm (grav)}\right]\right)$ 
(left top($+$)/bottom($-$)
 (Equations (\ref{quad},\ref{grav})), and the right-hand-side 
panels are for the magnetic part, 
namely $\log$ $[\left(\pm{A_{20}^{\rm{E} 2}}_{\rm (mag)}\right)]$ 
(right top($+$)/bottom($-$)) (e.g., Equation (\ref{mag})).
 By comparing the top two panels,
 it can be seen that the positive contribution is  
 overlapped with the regions where the MHD outflows exist.
The major positive contribution is from the kinetic term of the MHD outflows 
 with large radial velocities (e.g., $ + \rho_{*}W^2 {v_r}^2$  in Equation (\ref{quad})).
 The magnetic part also contributes to the positive trend (see top right-half
 in the right panel (labeled by mag(+))). This comes 
from the toroidal magnetic fields (e.g., $ + {b_{\phi}}^{2}$ in Equation
 (\ref{mag})), which dominantly contribute to drive MHD explosions. 

 Unfortunately, our MHD code becomes numerically unstable when the strong MHD jets 
 propagate to a stellar mantle with decreasing density, which prevents us from studying 
 the resulting GWs in more long run. 
The neutrino GWs are much smaller than the other GW sources (e.g., Figure \ref{f10}))
 simply due to the shorter simulation time. 
We expect that the increasing trends by the magnetic fields maintain as the 
 MHD shocks propagate further out (e.g., \citet{taki_kota}). To confirm it,
 we need to implement a numerical technique specially developed to solve 
the force-free fields, which is major undertaking (e.g., \citet{mckinney}).

\begin{figure}[htbp]
\begin{center}
\includegraphics[scale=0.6]{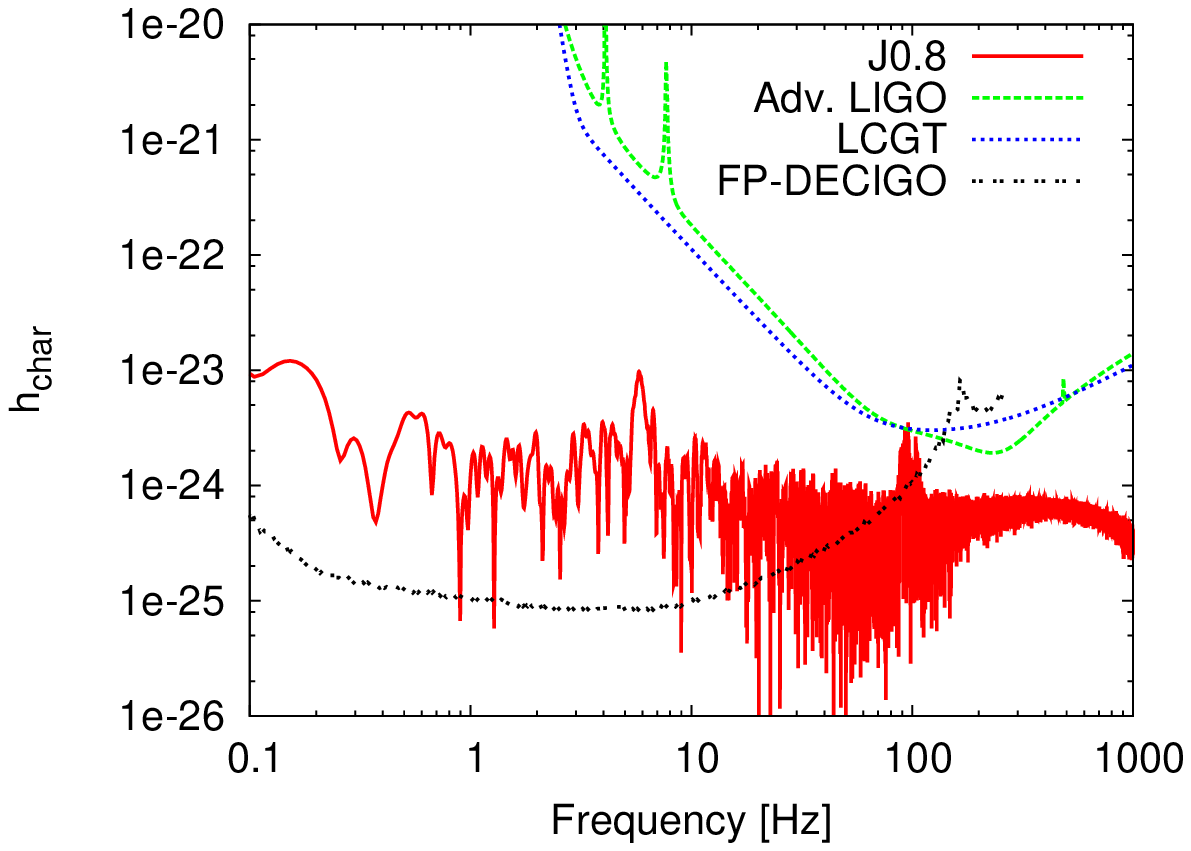}
\includegraphics[scale=0.6]{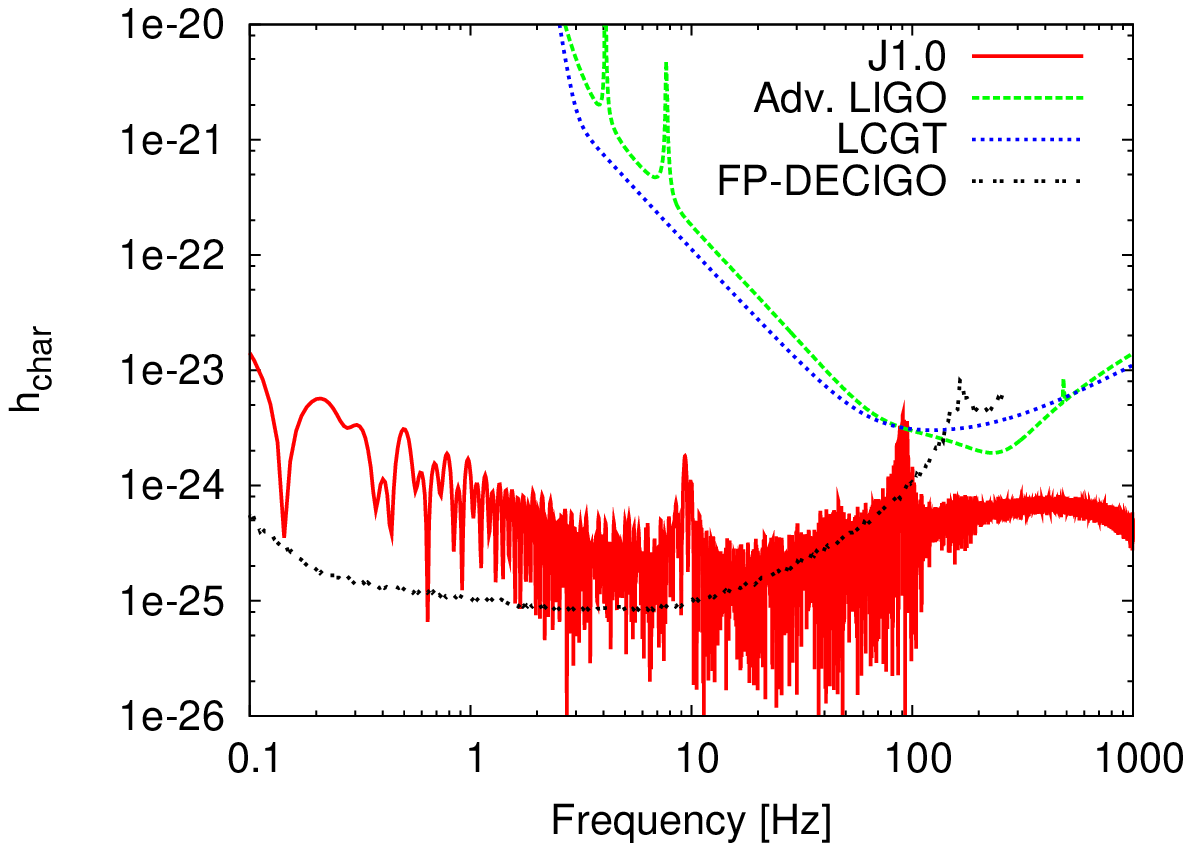}
\caption{Spectral distributions of GWs from matter motions for models J0.8 (left)
 and J1.0 (right) with 
the expected detection limits of advanced LIGO \citep{advancedligo}, Large-scale
 Cryogenic Gravitational wave Telescope (LCGT) \citep{lcgt}, and 
Fabry-Perot type DECIGO \citep{fpdecigo,kudoh}. The distance to the source
 is assumed to be 100 Mpc. $h_{\rm char}$ is
 the characteristic gravitational wave strain defined in \citet{flanagan}. 
Note that
 the frequency domain larger
 than $0.1$ Hz is only plotted, since the typical
 timescale that our simulations covered is 10 s.}
\label{f14}
\end{center}
\end{figure}

\subsection{Detectability}\label{3.4}

 Finally, Figure \ref{f14} depicts the GW spectra for models J0.8 (left) and J1.0 (right). 
 The GW spectra in the frequency domain between 1 to 100 Hz 
becomes slightly larger for model J0.8 than for model J1.0. 
This reflects a more efficient release of the gravitational binding energy 
for the moderately rotating case (model J0.8) as already mentioned 
 in sections 3.2 and 3.4.
  For the cosmological distance scale of GRBs ($\sim 100$ Mpc),
 these low frequency GW signals are unfortunately very hard to 
detect even by the advanced detectors (like the advanced LIGO or 
 KAGRA/LCGT) whose sensitivity is severely limited by the seismic noises 
\citep{tamanew,firstligonew,advancedligo,lcgt}. 
 A good news is that these signals could be detectable 
 by a recently proposed future space-based interferometer like Fabry-Perot type DECIGO 
(\citet{fpdecigo}, black line in Figure \ref{f14}). 
Two low-luminosity LGRBs were
 already observed at distances of $\sim 40$ Mpc (980425, \citet{gala98}) 
and $\sim 130$ Mpc (060218, \citet{fer06}). And their local rate, being 
 much higher than that of normal bursts, is expected to be as large 
$\approx 0.1 D_{100}~{\rm yr}^{-1}$ with $D_{100}$ 
representing the distance normalized by
 100 Mpc (see discussions in \citet{corsi09}).
 Our results suggest that the GW astronomy of
 collapsars could become reality by the DECIGO-class GW detectors, hopefully 
near in the future. 

\clearpage 

\section{Summary and Discussion}\label{summary}
By performing axisymmetric SRMHD simulations, we investigated
 possible signatures of GWs in the context of the collapsar model of LGRBs.
 By cutting out the central BH, we focused on the GWs generated by 
 asphericities in neutrino
 emission and matter motions in the vicinity of the hyperaccreting disks.
 Nine models were computed by changing initial angular momenta and magnetic fields parametrically in the precollapse core of a $35 M_{\odot}$ progenitor star.
 As for the microphysics,
 a realistic equation of state was employed and the 
neutrino cooling was taken into account via a multiflavor neutrino leakage scheme.
 To accurately estimate GWs from neutrinos, we performed a ray-tracing analysis in GR
 by a post-processing procedure.
 We studied also the effects of magnetic fields on the gravitational waveforms
  by employing a stress formula that includes contributions both from magnetic fields and special relativistic corrections. 
 We found that the GW amplitudes from
 anisotropic neutrino emission shows a monotonic increase with time, whose 
 amplitudes are much larger than those from matter motions of the accreting material.
 We showed that the increasing trend of the neutrino GWs stems from 
 the excess of neutrino emission in the direction near parallel to 
 the spin axis illuminated from the hyperaccreting disks. 
 We pointed out that a recently 
proposed future space interferometer like Fabry-Perot type DECIGO would permit the 
detection of these signals within $\approx$ 100 Mpc.

Here it should be noted that 
our 2D simulations cannot capture any non-axisymmetric instabilities so 
 far proposed to provide a strong GW emission in the semi-analytical 
(e.g., \citet{putten01,davies02,fryer02,kobayashi03,piro07,corsi09})
 and 
 in full GR simulations (e.g., \citet{shib05,manka07}). Therefore
   the present results might be regarded to give a lower limit for the possible 
 GW emission in collapsars. To go up the ladders beyond the 2D simulations
 is very numerical challenging, however, we need to handle it not only
 to test the outcomes of the proposed ideas about the non-axisymmetric instabilities
 but also to obtain more accurate waveforms from collapsars.

When studying the formation of BHs and the associated GW signals, 
the use of a pseudo-newtonian potential can lead to significant errors. 
This is the reason why 
 we had to limit our discussion only to the asphericities and the resulting 
 GW emission in the vicinity of the accretion disk which is far away from the 
 central object. In studying the dynamics from core-collapse to neutron star,
 a conformally-flat-condition (CFC) approximation has been often employed to solve 
 the GR equations \citep{dimm02,dimmelprl,dimm08}. 
And it has been tested that such a treatment is very good 
 in capturing the results of full-GR results (e.g., \citet{shib03})
 for a wide variety of supernova progenitors with neutron star formations. 
But it may need a further investigation
 that the approximation is still valid in applying to collapsars which 
 is a highly aspherical disk-BH system with very dilute polar funnel regions 
 along the rotational axis. Full GR simulations 
(e.g., \citet{baiotti07,ott2011,seki11,kuroda12} and 
 references therein) are indeed one of the most important topics pointing to 
the final frontiers of stellar core-collapse simulations, however, it is generally
 computationally too expensive at present to follow the late-time collapsar evolution
 up to $\sim $10 s, which is a typical duration of long-duration GRBs. Moreover
 the inclusion of microphysics such as neutrino heating is a major undertaking for 
 the full GR simulations. The Cowling approximation (or the fixed metric approach,
 e.g., \citet{mcki07b,komi07a,bark08}) has been often used in collapsar simulations 
 so far.
 But it still remains a non-trivial issue how to treat the self-gravity of the 
accretion disk. Finally, the pseudo-newtonian approach
 we take in this work cannot unambiguously capture accurate properties of the 
flows in the vicinity of the BH as well as the associated GW signals.
 Sacrificing the central regions, such simplified method would be currently only a 
 possible way to follow a long-term evolution (especially for 
  the disk evolution) with including an appropriate treatment of microphysics.
 Needless to say, these four approaches (full GR, CFC, Cowling, and post-newtonian) 
may be regarded to be complimentary useful to study the different epochs in 
 the collapsar evolution. For bridging the gaps between them, it may be a good 
idea to employ the end results of fully GR simulations (e.g., \citet{ott2011}) in our 
 simulation, which we consider to be 
 the most urgent task to investigate as a sequel of this study.

 In a data analysis to extract the true GW signals from
 the confusing detector noises, it should be of 
primary importance to take a coincident analysis with the conventional electromagnetic 
observation as well as neutrinos. What could be the photon and neutrino signatures 
in our collapsar models ? To answer this question, we need to perform a 
 long-term simulation that bridges continuously the phase 2 and 3 as 
mentioned in the introduction. For generating neutrino-driven or MHD-driven outflows 
in numerical simulations of collapsars, the GRMHD simulations including 
the effects of the neutrino heating are needed, the formulation of which is in a 
 steady progress (e.g., \citet{shibata11,mueller,kuroda12}). Assisted by a growing 
computational power, these advanced simulations would be 
hopefully practicable by utilizing the next-generation supercomputers.
 These updates should bring forward not only our understanding the dynamics of 
 the collapsar engines, but also the theoretical predictions of
 observable multi-messengers (e.g., \citet{ando12} for a recent review\footnote{see,
 \citet{kotake12} for a review about multi-messenger perspectives on core-collapse
 supernovae.}), 
including 
neutrino emission (e.g., \citet{icecube}) and nucleosynthetic yields
 (e.g., \citet{fujimoto07} and references therein). The neutrino signals 
 emitted from our collapsar models (e.g., from Figure \ref{f2} with its typical 
 emergent neutrino energy of $\sim 15 - 20$ MeV) are visible up 
 to the Local Group ($\sim 1$ Mpc, Kawagoe et al. in preparation) 
by future megaton-class detectors (e.g., Hyper-Kamiokande, Memphys, and LBNE), 
  large-scale scintillators (e.g., HALO \citep{engel}, and by liquid-Argon detectors
 (GLACIER, see \citet{scholberg} for collective references therein).
  While this work might raise many more questions than it can
 answer, it definitely makes clear that our understanding of the 
GWs in collapsars is still in its infancy and that collapsars and the BH-forming 
 supernovae are "gold mine" in which a number of unsettled and fascinating 
 research themes are hidden. 
We hope that our exploratory results, at least, give momentum to 
theorists to make the GW prediction based on a more sophisticated 
 numerical modeling of collapsars.


\acknowledgements{K.K. and T.T. are thankful to K. Sato and S. Yamada for continuing 
encouragements. We highly appreciate to our anonymous referee for our careful 
 reading of our manuscript with valuable comments. 
Numerical computations were carried out in part on XT4 and 
general common use computer system at the center for Computational Astrophysics, CfCA, 
the National Astronomical Observatory of Japan.  This 
study was supported in part by the Grants-in-Aid for the Scientific Research 
from the Ministry of Education, Science and Culture of Japan (Nos. 19540309, 20740150,
 23540323, and 23340069) and by HPCI Strategic Program of Japanese MEXT.}

\end{document}